\documentclass{article}
\usepackage{amsmath,amssymb}
\usepackage{cite}
\usepackage{geometry}
\usepackage{hyperref}
\usepackage{url,slashed}
\usepackage{amsmath,amssymb}   
\usepackage{epsfig}                    
\usepackage[matrix,arrow]{xy}          
\usepackage{xspace}                    
\usepackage{stmaryrd}                  
\usepackage{slashed}
\usepackage{appendix,graphicx}
\usepackage{enumitem}
\usepackage{cite}
\usepackage{hyperref}
\usepackage{multirow}

\usepackage{units} 
\usepackage{tikz}
\usepackage{color}

\usepackage[vcentermath]{youngtab}

\newcommand\fft[2]{\frac{#1}{#2}}
\newcommand\ft[2]{{\textstyle\frac{#1}{#2}}}

\newcommand\tr{\operatorname{tr}}
\mathchardef\mhyphen="2D

\newcommand{\JJ}{\mathcal {J}}

\def\cL{\mathcal{L}}

\def\cO{{{\mathcal O}}}
 
\def\cN{\mathcal{N}}

\def\bZ{\mathbb{Z}}

\def\pa{\partial}
\def\na{\nabla}
\def\fr{\frac}

\def\ra{\rightarrow}

\def\a{\alpha}
\def\b{\beta}

\def\e{\epsilon}

\def\l{\lambda}

\def\m{\mu}
\def\n{\nu}

\def\O{\Omega}

\def\r{\rho}
\def\s{\sigma}

\def\t{\tau}

\def\z{\zeta}

\newcommand\beq{\begin{equation}}
\newcommand\eeq{\end{equation}}
\newcommand\bea{\begin{eqnarray}}
\newcommand\eea{\end{eqnarray}}
\newcommand\nn{\nonumber}

\makeatletter
\@addtoreset{equation}{section}
\makeatother

\begin{document}
\setcounter{page}{0}
\begin{titlepage}
\titlepage

\begin{flushright}
IPhT-t19/168 \qquad LCTP-19-36
\end{flushright}

\vspace{30pt}
\begin{center}

{\Large {\bf Higher-derivative couplings in string theory:  \\
\vspace{10pt}  five-point contact terms}}

\vspace{25pt}

James T. Liu$^a$ and Ruben Minasian$^b$

\vspace{20pt}

{${}^a$\it Leinweber Center for Theoretical Physics\\
Randall Laboratory of Physics, The University of Michigan\\
Ann Arbor, MI 48109--1040, USA}

\vspace{10pt}

{${}^b$\it Institut de Physique Th\'eorique, Universit\'e Paris Saclay, CNRS, CEA\\
91191 Gif-sur-Yvette Cedex, France}

\vspace{40pt}

\underline{ABSTRACT}
\end{center}

We compute the tree-level $H^2R^3$ couplings of type II strings and provide some basic tests of the couplings by considering both K3 and Calabi-Yau threefold compactifications.  Curiously, additional kinematical structures show up at tree level that are not present in the one-loop couplings.  This has interesting implications for type II supersymmetry as well as $SL(2, \mathbb Z)$ duality in type IIB strings.

\vfill
\begin{flushleft}
December 2019
\vspace{.5cm}
\end{flushleft}
\end{titlepage}

\newpage

\tableofcontents

\section{Introduction}

In ten-dimensional theories with 32 supercharges, $\alpha'$ corrections start at eight-derivative level and receive perturbative contributions at tree level \cite{Schwarz:1982jn,Gross:1986iv,Grisaru:1986px,Grisaru:1986dk,Grisaru:1986kw,Freeman:1986br,Grisaru:1986vi,Freeman:1986zh} and at one loop \cite{Sakai:1986bi}. These contributions begin from four-field scattering, where the four-graviton part is the most studied and best understood one, hence the moniker `$R^4$ couplings'. In the string frame, the entire NS sector results are compactly written using \cite{Gross:1986mw}
\begin{equation}
\label{eq:lin}
\tilde{R}_{\mu\nu}{}^{\alpha\beta}  = (R^{\, \rm lin})_{\mu\nu}{}^{\alpha\beta} +\nabla_{[\mu}H_{\nu]}{}^{\alpha\beta} \, ,
\end{equation}
where $(R^{\, \rm lin})_{\mu\nu \r \l} =-\fft12( \partial_{\m} \partial_{\r} h_{\n \l} + \cdots) = - 2 \partial_{[\m}h_{\n][\r,\l]}$ is the linearised Riemann tensor.  The tree-level corrections are given by
\begin{equation}
\label{eq:tree}
e^{-1}\mathcal L\sim e^{-2\phi}(t_8t_8 \tilde{R}^4-\ft14\epsilon_{8}\epsilon_{8} \tilde{R}^4) , 
\end{equation}
where the first pair of indices on each $\tilde R$ is contracted on the first $t_8$ or $\epsilon_{8}$, and the second pair on the second \cite{Gross:1986iv}.  Note that the original calculations were performed in the Green-Schwarz formalism so $\epsilon_8$ is the fully antisymmetric tensor in the eight-dimensional transverse Euclidean space%
\footnote{To be precise, the four-point function in light-cone gauge is not sensitive to the $\epsilon_8\epsilon_8$ term.  However, the structure of the amplitude indicates that it is present.}.
However, we take a more covariant approach and regard $\epsilon_n\epsilon_n$ as a shorthand notation for the anti-symmetric delta function on $2n$ indices with a precise definition given below in (\ref{eq:epsepsdef}).  The tensor $t_8$ has four pairs of antisymmetric indices, and is such that given an antisymmetric matrix $M$, $t_8 M^4 =  24 \left(\tr M^4 - \ft14 (\tr M^2)^2\right)$. The CP-even sector of the one-loop expression has a similar structure:
\begin{equation}
\label{eq:one}
e^{-1}\mathcal L_{\rm CP\mhyphen even}\sim (t_8t_8 \tilde{R}^4\pm\ft14\epsilon_{8}\epsilon_{8}\tilde{R}^4),
\end{equation}
where the top (bottom) sign is for the IIA (IIB) theory.  The two terms here come respectively from the even-even and odd-odd spin structure sectors of the covariant one-loop amplitude.

Due to general covariance, the purely gravitational part of the higher-derivative couplings can be completed to expressions involving the full Riemann tensor without needing to compute higher-point graviton amplitudes%
\footnote{The exception is for the couplings that vanish on shell at linearised level. This is the case for e.g.\ the $\epsilon_{8}\epsilon_{8}\tilde{R}^4$ terms in \eqref{eq:tree} and \eqref{eq:one}). These couplings were first obtained using duality arguments, rather than five-point function calculations.}.
Hence, the gravitational part of the eight-derivative couplings is well-known. However, the full completion including the anti-symmetric tensor and possibly dilaton is not yet known.
While it is tempting to introduce a four-index tensor computed using a connection with torsion 
\begin{equation}
 R_{\mu\nu}{}^{\alpha\beta}(\Omega_+)\equiv
R_{\mu\nu}{}^{\alpha\beta}
+\nabla_{[\mu}H_{\nu]}{}^{\alpha\beta}+\ft12H_{[\mu}{}^{\alpha\gamma}H_{\nu]\gamma}{}^\beta
\end{equation}
with $(\O_\pm)_{\m}{}^{\a\b}  = \Omega_{\m}{}^{\a\b} \pm \fr12 H_{\m}{}^{\a\b}$ as a natural non-linear counterpart to $\tilde R$ in \eqref{eq:lin}, there is no a priori reason to believe that such a replacement can capture the complete string-theoretic answer%
\footnote{For a closed $H$, $R_{\m \n\r\l}(\Omega_+) = R_{\r\l\m\n}(\Omega_-)$. Clearly,  $R_{\m \n\r\l}(\Omega_+)$ is antisymmetric in the first and the second pairs of indices. Moreover, $t_8t_8 {R}(\O_+)^4 = t_8t_8 {R}(\O_-)^4$ and $\epsilon_{8}\epsilon_{8} R(\O_+)^4 = \epsilon_{8}\epsilon_{8} R(\O_-)^4$. }.
In fact, it is known that additional kinematic structures beyond the
standard $\epsilon \epsilon R^4(\Omega_+)$ appear in terms involving higher powers of $H$, such as
$H^2 R^3$ and $H^4 R^2$ in the odd-odd sector of the one-loop $R^4$ couplings. These have been verified by five \cite{Peeters:2001ub,Richards:2008sa}  and partial six-point function calculations \cite{Liu:2013dna}. However no tree-level results are known at the level of the effective action beyond four-point functions. Computation of the tree-level $H^2 R^3$ couplings is the main focus of this paper.

One may suspect that we have quite a bit of indirect information about these terms. After all the ten-dimensional $R^4$ couplings are at the origin of the perturbative corrections to the four-dimensional ${\mathcal N}=2$ moduli spaces as well as the perturbative $R^2$ couplings in four-dimensional ${\mathcal N}=2$ and ${\mathcal N}=4$ theories. However the lower-dimensional results do not rely on reduction of the ten-dimensional couplings. This might be best illustrated by looking at $R^2$ terms in type II compactifications on $K3$. There supersymmetry arguments are strong enough to rule out such couplings at tree-level for either IIA or IIB.  In type IIB compactified on $K3$ the $R^2$ couplings at one loop are also absent. At   the linearised level it is not hard to find cancellations in the reductions of the $R^4$ couplings that assure the absence of these terms, using $R_{\mu \nu} = R = 0$. Yet the complete cancellation at the non-linear level has not been checked, and as we shall see is rather nontrivial. The same is true for the  $R^2$ terms in type II compactifications on Calabi-Yau three-folds, where special geometry dictates the absence of $R^2$ terms in IIB compactifications and the absence of tree-level $R^2$ terms for IIA.

Recently  a new $H^2R^3$ term that does not follow the standard tree-level kinematics was proposed in \cite{Grimm:2017okk} in order to make $R^4$ reductions compatible with the quantum corrections to the moduli space metrics for four-dimensional ${\mathcal N}=2$  theories  \cite{Antoniadis:1997eg, Antoniadis:2003sw}. We confirm this proposal and find many more couplings that cannot be detected by examination of two or four-derivative actions in compactifications without fluxes on four- or six-dimensional Ricci flat spaces. Our tree-level result can be summarised as:
\begin{align}
\mathcal L_{\mathrm{tree}}&=\sqrt{-g}\,e^{-2\phi}\biggl[\fft{\zeta(3)}{3\cdot2^{11}}\alpha'^3
\Bigl(t_8t_8R(\Omega_+)^4-\ft14\epsilon_{8}\epsilon_{8} R(\Omega_+)^4-2t_8t_8H^2R(\Omega_+)^3-\ft16\epsilon_{9}\epsilon_{9}H^2R(\Omega_+)^3\nn\\
&\kern10em+8\cdot4!\sum_id_iH^{\mu\nu\lambda}H^{\rho\sigma\zeta}\tilde Q^i_{\mu\nu\lambda\rho\sigma\zeta}+\cdots\Bigr)+\cdots\biggr].
\label{eq:treeea**}
\end{align}
Here only the first two terms have the standard kinematics, exactly as in \eqref{eq:tree}, with the curvature tensor now computed using the connection with torsion. The last two terms on the first line use the familiar tensorial structures $t_8$ and $\epsilon_{10}$ but the structure of indices on the $H^2$ part is such that it cannot be obtained from expanding the standard terms. Their explicit form is as follows:
\begin{align}
t_8t_8H^2R(\Omega_+)^3&\equiv t_{8\,\mu_1\cdots\mu_8}t_8^{\nu_1\cdots\nu_8}H^{\mu_1\mu_2\alpha}H_{\nu_1\nu_2\alpha}R^{\mu_3\mu_4}{}_{\nu_3\nu_4}(\Omega_+)R^{\mu_5\mu_6}{}_{\nu_5\nu_6}(\Omega_+)R^{\mu_7\mu_8}{}_{\nu_7\nu_8}(\Omega_+),\nn \\
\epsilon_{9}\epsilon_{9}H^2R(\Omega_+)^3&\equiv-\epsilon_{\alpha\mu_0\mu_1\cdots\mu_8}\epsilon^{\alpha\nu_0\nu_1\cdots\nu_8}
H^{\mu_1\mu_2}{}_{\nu_0}H_{\nu_1\nu_2}{}^{\mu_0}
R^{\mu_3\mu_4}{}_{\nu_3\nu_4}(\Omega_+)
R^{\mu_5\mu_6}{}_{\nu_5\nu_6}(\Omega_+)R^{\mu_7\mu_8}{}_{\nu_7\nu_8}(\Omega_+).
\label{eq:eeH2R3**}
\end{align}
These terms are crucial for finding agreement with the known lower-dimensional results, notably the corrections to the moduli space metrics of ${\mathcal N} =2$ vector and hypermultiplets. The $\epsilon_{9}\epsilon_{9}H^2R(\Omega_+)^3$ kinematical structure also appears at one loop, albeit with a different coefficient. (The one-loop result is given in \eqref{eq:loopea}.) The $t_8t_8H^2R^3$ term appears only in the tree-level expressions%
\footnote{Even though couplings quadratic in $H$ and cubic in curvature appear in the expansion of the terms with the standard kinematics, we shall be reserving the notation $t_8t_8H^2R^3$ and $\epsilon_{9}\epsilon_{9}H^2R^3$ exclusively for the couplings defined in \eqref{eq:eeH2R3**}.  Note that the definitions $\Delta J_0(\Omega_+, H)\equiv-\epsilon_{9}\epsilon_{9}H^2R^3$ was used in  \cite{Liu:2013dna}, and $\delta {\mathcal J}\equiv-2 t_8t_8H^2R^3$ was used in \cite{Grimm:2017okk}.}.

The last line of \eqref{eq:treeea**} contains only terms without any index contractions between the two $H$'s. Here  $d_i$ is a set of constants  given in \eqref{eq:di} and the quantities $Q^i_{\mu\nu\lambda\rho\sigma\zeta}$ are cubic in Riemann curvatures and are specified in \eqref{eq:Qi}. The appearance of such terms does come as a bit of a surprise, as we have not been able to rewrite them in terms of $t_8$ and $\epsilon_{10}$, and they do not have any one-loop counterparts.  Note, however, that they are undetectable from the lower-dimensional physics obtained via compactification without fluxes.

The couplings appearing in non-linear completions are of two types: those which fit the kinematic structure that appears by substituting $R \, \ra \, R(\Omega_+)$ in the $R^4$ couplings, (\ref{eq:tree}) and (\ref{eq:one}), and those which do not.  At one loop, to the highest order we have checked so far (partially up to six point at one-loop), we indeed find a combination of `standard' terms coming from the expansion of $t_8t_8 R(\O_+)^4$ and $\epsilon_8\epsilon_8R(\O_+)^4$ as well as terms of the second type involving a pair of $H$-fields not contracted with each other that are of the form  $\epsilon_9\epsilon_9 H^2 X $, where $X \sim R^3$ or $X \sim (\nabla H)^2 R$ (see \eqref{eq:eeH2R3} and \eqref{eq:eeH2NH2R} for the explicit form). We find a much more elaborate structure at tree level.  As mentioned, our partial knowledge of the couplings of the second type have been mostly indirect; the lower-dimensional implications of these terms have been deduced by methods other than the reduction of ten-dimensional couplings. The completion with the dilaton and with the RR fields is still not known. We have collected the results of known direct string-theoretic calculations in Table \ref{Tab:whatweknow}.

\begin{table}[t]
\label{Tab:whatweknow}
\begin{center}
\begin{tabular}{|c|c|c|c|c|c|c|c|c|c|c|}
\hline
\multicolumn{2}{|c|} {\centering } & \multicolumn{3}{|c|} {\centering 4pt} & \multicolumn{3}{|c|} {\centering 5pt} & \multicolumn{3}{|c|} {\centering 6pt} \\ \hline
\multicolumn{2}{|c|} {\centering } & $g, B$ & $\phi$ & RR & $g, B$ & $\phi$  & RR & $g, B$ & $\phi$  & RR  \\ \hline
\multicolumn{2}{|c|} {\multirow{4}{*}{\centering tree}} &\multirow{4}{*} {$\varnothing$} &\multirow{4}{*}  {$\varnothing$} & \multirow{4}{*} {\cite{Policastro:2006vt,Policastro:2008hg} }&  $-\ft16\epsilon_{9}\epsilon_{9}H^2R^3$ & $\multirow{4}{*} {?}$ & $\multirow{4}{*} {?}$  & $\multirow{4}{*} {?}$ & $\multirow{4}{*} {?}$  & $\multirow{4}{*} {?}$ \\
\multicolumn{2}{|c|} {\centering } & &  && $-2t_8t_8H^2R^3$ & & & &   &  \\
\multicolumn{2}{|c|} {\centering } & &  &&  $ 8\cdot 4! \sum d_i H^2 \cdot {\tilde Q}^i $ & & &  & &   \\ 
\multicolumn{2}{|c|} {\centering } & &  &&  $\sim H^2 (\nabla H)^2 R \,\, ?$ & & &  & &   \\ \hline
\multirow{3}{*}{1-loop}& \multirow{2}{*}{\centering o-o}&  \multirow{2}{*}{$\varnothing$} & \multirow{2}{*}{$\varnothing$} & $\multirow{2}{*}{?}$ & $ \pm\fft13 \epsilon_{9}\epsilon_{9}H^2R^3$ & \multirow{2}{*}{$\varnothing$} & $\multirow{2}{*}{?}$  &\multirow{2}{*}{ partial   \cite{Liu:2013dna}}& \multirow{2}{*}{$\varnothing$}  & $\multirow{2}{*}{?}$ \\
& & \multicolumn{1}{c|}{} & \multicolumn{1}{c|}{ } & & $\mp \fft{4}{9}  \epsilon_{9}\epsilon_{9}H^2 (\nabla H)^2 R$  & &  &  &   &   \\
\cline{2-11} & \multicolumn{1}{c|}{e-e} & \multicolumn{1}{c|}{$\varnothing$} & \multicolumn{1}{c|}{$\varnothing$} & $?$ &{$\varnothing$}&{$\varnothing$}& $?$ & $\varnothing$ & $\varnothing$ & $?$ \\ \hline
\end{tabular}
\end{center}
\caption{Summary of additional contributions to the effective action in string frame which complete the $R^4$ terms beyond $R \, \ra \, R(\Omega_+)$ known from the string-theoretic calculations only. Separate columns denote the knowledge about the dilaton and RR couplings. The verified absence of couplings involving the given field is denoted by $\varnothing$. Our ignorance is denoted by $?$.  The NSNS contribution to the CP-odd part (even-odd and odd-even) of the one-loop couplings is completely determined by  $R \, \ra \, R(\Omega_+)$, and does not involve the dilaton. Its RR completions is not known. In this paper we compute the $\sim H^2R^3$ contributions at tree-level. We have not  done the computation for  the terms $\sim  H^2 (\nabla H)^2R$ which could also come from five NSNS field scattering. The one-loop $ \epsilon_{9}\epsilon_{9}  H^2(\nabla H)^2R$ structure is defined in \eqref{eq:eeH2NH2R}. }
\end{table}

It should be noted that much more is known at the level of string amplitudes.  Three and four point functions were naturally investigated as part of the development of the superstring formalism.  Shortly thereafter, the tree-level open string five-point amplitude was computed in \cite{Kitazawa:1987xj}, and the one-loop amplitude was computed in \cite{Frampton:1985uw,Frampton:1986ea,Frampton:1986gi,Lam:1986kg}.  Closed string amplitudes can of course be obtained from the open string ones using the KLT relations \cite{Kawai:1985xq} that were also developed early on.  Closer to the present, the pure-spinor formalism \cite{Berkovits:2000fe} provided a fruitful alternative to the traditional covariant and Green-Schwarz approaches to string amplitudes \cite{Mafra:2009wq,Mafra:2011nv,Mafra:2011nw,Mafra:2012kh}.  This formalism also allows for a unified treatment of the NS and R sectors, thus making the scattering of RR fields more feasible and was used in \cite{Policastro:2006vt,Policastro:2008hg} to obtain the complete (to all orders of $\alpha'$) tree-level quartic effective action of the type II string.  The challenge we face is not in computing the higher-point closed string amplitudes, but rather in constructing a local eight-derivative effective action that reproduces these amplitudes.  In addition to the large number of kinematical structures that show up, we are also faced with the issue of subtracting out pole contributions from lower-point amplitudes in order to recover only the new contact terms that show up at each higher order in the expansion.

In a search for a (generalised) geometric description of string theory, the rather baroque structures appearing at the non-linear (beyond four-point) level present an interesting challenge. In this regard, the torsionful connection and $R(\O_+)$ is very natural. Yet clearly between the tree-level and one-loop terms already at the first nonlinear level (five-point functions) $H^2$ terms also enter independently from the $R \, \ra \, R(\Omega_+)$ substitution, and come in diverse kinematical structures. As mentioned, writing the tree-level five-point function contributions quadratic in $H$ using the familiar $t_8$  or $\epsilon_{10}$ contractions (or generalisations) does not appear to be possible. A somewhat loose analogy might be found in eight-derivative couplings involving the graviton and the axi-dilaton in type IIB, where already at the four-point level the kinematics is rather unwieldy \cite{Policastro:2008hg}. Yet in an F-theoretic context one can see the emergence of these structures from purely gravitational couplings in twelve-dimensional elliptically fibered spaces \cite{Minasian:2015bxa}.  Finding a similar geometrisation of non-linear completions of the $R^4$ couplings (and beyond) would be of great importance.

From the low-energy point of view this mismatch between tree and one-loop terms might appear somewhat surprising.  For example, in type IIB the linearised terms are identical at tree-level and one-loop. As mentioned, the six-dimensional $(2,0)$ theory is not supposed to have $\sim R^2$ couplings either at tree-level or at one-loop. And indeed it is not hard to check that at this order the different contributions coming from $R^4$ terms cancel when compactified on $K3$. Denoting the internal and six-dimensional curvatures by $R_0$ and $\tilde R$ respectively, one finds that on $K3$,  $t_8 t_8 R^4 \, \mapsto \, t_4 t_4 R_0^2 \times t_4 t_4 {\tilde R}^2$ and 
$\e_{8} \e_{8} R^4 \, \mapsto \, \e_4 \e_4 R_0^2 \times \e_4 \e_4 {\tilde R}^2$.  Since the internal parts are the same due to Ricci-flatness, the IIB combination $(t_8t_8-\fft14\epsilon_8\epsilon_8)R^4$ reduces to $(t_4 t_4- \frac14 \e_4 \e_4)\tilde R^2 = 4\tilde R_{\m \n} R^{\m \n}-\tilde R^2$, which contains only Ricci terms and vanishes on-shell at the linearised level.  This works exactly the same way at tree-level and at one-loop. However, as we shall see in section \ref{sec:K3}, the complete non-linear cancellations at tree-level and at one-loop are due to very different mechanisms.

The mismatch between the tree-level and one-loop terms in the non-linear higher-derivative couplings (starting from five-field scattering) also poses questions for supersymmetry and $SL(2,\mathbb Z)$ invariance of the type II theory.  Although IIA and IIB theories are distinct, the eight-derivative supersymmetry invariants are often given in terms of $\mathcal N=1$ combinations.  At the linearised level, it is generally believed that there are only two $\mathcal N=1$ superinvariants, given schematically as
\begin{align} \label{eq:superinv}
J_0&=(t_8t_8-\ft14\epsilon_8\epsilon_8)R^4,\nn\\
J_1&=t_8t_8R^4-\ft14\epsilon_{10}t_8BR^4.
\end{align}
Here $J_0$ gives the IIA and IIB tree-level couplings as well as the IIB one-loop coupling, while the combination $2J_1-J_0$ gives the IIA one-loop coupling.  However, this cannot hold at the non-linear level since the tree-level structure is distinct from that of any of the one-loop invariants.  We suggest that \eqref{eq:superinv}, suitably completed by nonlinear terms, become the one-loop superinvariants (see \eqref{superinv1}) and that there must be at least one additional tree-level superinvariant, which can be viewed as the completion of $t_8t_8e^{-2\phi}R^4$.  The tree-level dilaton factor cannot be probed by a four-point function, but will affect the supersymmetry completion as the dilaton, being in the supergravity multiplet, transforms non-trivially under supersymmetry.

For the IIB string, there is also the issue of $SL(2,\mathbb Z)$ invariance of the eight-derivative couplings to consider.  In the purely gravitational sector, the $J_0$ combination given above is complete even at the nonlinear level, and the relative tree and loop factors multiplying $J_0$ give rise to the $SL(2,\mathbb Z)$ invariant \cite{Green:1997tv, Green:1997di}
\begin{equation}
\mathcal L_{IIB}^{\partial^8}\sim\sqrt{-g}\mathcal E_{3/2}(\tau,\bar\tau)J_0,
\end{equation}
where $\mathcal E_{3/2}(\tau,\bar\tau)$ is a non-holomorphic Eisenstein series of weight $3/2$.  However, once this is expanded to the full supergravity multiplet, additional terms beyond $J_0$ will enter, and moreover will transform with different modular weights.  In this case, the story of $SL(2,\mathbb Z)$ invariance becomes much more intricate, and additional knowledge of the RR sector will be needed to complete the picture.  Nevertheless, we will show that in some cases $SL(2,\mathbb Z)$ invariance can be used to extend the NSNS sector couplings to the complete set of IIB fields. 

In fact, $SL(2,\mathbb Z)$ invariance of the type IIB string gives rise to some tension between the different four-point results collected in Table~\ref{Tab:whatweknow}. For IIB strings, the only difference between the quartic NSNS contributions at tree level and one loop is the dilaton factor $e^{-2\phi}$.  Other than that, the kinematic structures are identical. This means that the only way of completing the purely NSNS expressions to $SL(2, \mathbb Z)$ invariant ones is by making each term invariant and multiplying the entire expression quartic in fields by the $SL(2, \mathbb Z)$ function $\mathcal E_{3/2}(\tau,\bar\tau)$. In particular this means that the local $U(1)$ symmetry of type IIB supergravity is respected by the four-particle interactions. This is indeed consistent with the results of \cite{Boels:2012zr, Green:2019rhz} showing that the $U(1)$-violating contributions start at the level of five-particle interactions. From the other side, the four-point result including RR fields given in \cite{Policastro:2008hg} cannot be completed to an $SL(2, \mathbb Z)$ invariant expression without modular forms that transform under weights $\pm 1$, and are hence $U(1)$-violating. We shall return to this issue in subsection \ref{SL(2,Z)}.

The structure of the paper is as follows. In section ~\ref{sec:stam} we first review the construction of the quartic effective action in the NSNS sector at tree-level and one-loop.  We then turn to the one-loop and tree-level five point function and the quintic effective action.  Although our main interest is in the tree-level effective action, we review the one-loop computation \cite{Liu:2013dna} as we use it as a reference in reconstructing the tree-level action.  In section~\ref{sec:K3}, we present two tests of the quintic effective action.  The first is the reduction on K3 and comparison to the known structure of six-dimensional $(1,1)$ and $(2,0)$ theories and the second is the reduction on Calabi-Yau threefolds.  We then turn to the question of $\mathcal N=1$ superinvariants followed by a discussion of $SL(2,\mathbb Z)$ invariance of the IIB couplings in section~\ref{sec:susy}.  Finally, we provide a brief summary of open issues in section~\ref{sec:disc}.

\section{The tree-level five-point function}
\label{sec:stam}

The eight-derivative terms start at the level of $\alpha'^3R^4$, corresponding to the scattering of four closed-string states.  Since the four-point string amplitude is needed for recreating the effective action, we begin with a brief review before turning to the five-point function.

\subsection{Open and closed string four-point functions}

While we are interested in closed string amplitudes, the KLT relations \cite{Kawai:1985xq} allow us to start with open string amplitudes as the basic building blocks.  The tree \cite{Green:1981xx} and one-loop \cite{Green:1981ya} open string four-point functions have been known since the introduction of the superstring, and both have the same kinematical form corresponding to the Yang-Mills four-point amplitude
\begin{equation}
A_{\mathrm{YM}}(1,2,3,4)=-\fft{t_8(k_1,e_1,k_2,e_2,k_3,e_3,k_4,e_4)}{k_1\cdot k_2 k_2\cdot k_3}.
\end{equation}
Here we have used the notation
\begin{equation}
t_8(\zeta_1,\zeta_2,\ldots,\zeta_8)=t_{8\,\mu_1\mu_2\cdots\mu_8}\zeta_1^{\mu_1}\zeta_2^{\mu_2}\cdots\zeta_8^{\mu_8},
\end{equation}
and the $t_8$ tensor is given by \cite{Schwarz:1982jn}
\begin{align}
t_{8\,\mu_1\nu_1\cdots\mu_4\nu_4}&=-2\bigl((\eta_{\nu_2\mu_1}\eta_{\nu_1\mu_2})(\eta_{\nu_4\mu_3}\eta_{\nu_3\mu_4})+(\eta_{\nu_3\mu_1}\eta_{\nu_1\mu_3})(\eta_{\nu_4\mu_2}\eta_{\nu_2\mu_4})+(\eta_{\nu_4\mu_1}\eta_{\nu_1\mu_4})(\eta_{\nu_3\mu_2}\eta_{\nu_2\mu_3})\bigr)\nn\\
&\quad+8\bigl(\eta_{\nu_4\mu_1}\eta_{\nu_1\mu_2}\eta_{\nu_2\mu_3}\eta_{\nu_3\mu_4}+\eta_{\nu_4\mu_1}\eta_{\nu_1\mu_3}\eta_{\nu_3\mu_2}\eta_{\nu_2\mu_4}+\eta_{\nu_2\mu_1}\eta_{\nu_1\mu_3}\eta_{\nu_3\mu_4}\eta_{\nu_4\mu_2}\bigr),
\end{align}
where the right-hand side is to be antisymmetrized in all $[\mu_i\nu_i]$ index pairs with weight one.  The external states are massless and on-shell, with momenta $k_i$ and polarizations $e_i$ satisfying $k^2=0$, $k\cdot e=0$ and momentum conservation, $k_1+k_2+k_3+k_4=0$.  The actual open-string amplitude is expanded in on-shell momenta.  At tree-level, the leading term is directly proportional to $A_{\mathrm{YM}}(1,2,3,4)$, and reproduces the four-point Yang-Mills contact interaction as well as $s$-, $t$- and $u$-channel gluon exchange diagrams, while the next term in the $\alpha'$ expansion gives a stringy four-derivative correction of the form $t_8F^4$.  At one-loop, the leading term starts at the four-derivative level, and has the same kinematical structure of $t_8F^4$.

As demonstrated in \cite{Kawai:1985xq}, closed string amplitudes can be written as a combination of left- and right-moving open-string amplitudes.  Working in the NSNS sector, we take the closed string polarization to be a tensor product
\begin{equation}
\theta_{\mu\nu}=e_\mu\otimes\bar e_\nu.
\end{equation}
This can be decomposed into a combination of NSNS fields according to
\begin{equation}
\theta_{\mu\nu}=h_{\mu\nu}+b_{\mu\nu}+\ft12(\eta_{\mu\nu}-\bar k_\mu k_\nu-k_\mu\bar k_\nu)\phi,
\label{eq:thetapol}
\end{equation}
where $h_{\mu\nu}$ are transverse-traceless metric fluctuations, $g_{\mu\nu}=\eta_{\mu\nu}+h_{\mu\nu}$, with $k^\mu h_{\mu\nu}=h^\mu_\mu=0$, and $b_{\mu\nu}$ are anti-symmetric tensor fluctuations with $H=db$ and $k^\mu b_{\mu\nu}=0$.  The dilaton $\phi$ corresponds to the trace mode, and $\bar k$ is introduced with $\bar k^2=0$ and $k\cdot\bar k=1$ in order to enforce the transversality condition $k_\mu\theta^{\mu\nu}=0$ for all modes including the dilaton.  While $\bar k$ is only implicitly defined by these properties, it drops out of all physical amplitudes involving the dilaton.

At tree-level, the closed string four-point function takes the form \cite{Gross:1986iv}
\begin{equation}
M_4^{\mathrm{tree}}\sim\left(\fft8{s_{12}s_{23}s_{13}}+\alpha'^32\zeta(3)+\cdots\right)\left|s_{12}s_{23}A_{\mathrm{YM}}(1,2,3,4)\right|^2,
\label{eq:M4t}
\end{equation}
where $s_{ij}=k_i\cdot k_j$.  This expression is valid for the scattering of any combination of gravitons, antisymmetric tensors and dilatons (although it vanishes by world-sheet parity for the scattering of an odd number of $b_{\mu\nu}$'s).  The first term reproduces the leading two-derivative action in the Einstein frame, while the second term gives rise to the familiar eight-derivative coupling of the form $t_8t_8R^4$.  In fact, $M_4^{\mathrm{tree}}$ provides information on the full closed-string NSNS sector, and it is easy to see that the eight-derivative coupling is built from the gauge invariant combination
\begin{equation}
\bar{\tilde R}^{\mu_1\mu_2}{}_{\nu_1\nu_2}=2\theta^{[\mu_1}{}_{[\nu_1}k^{\mu_2]}k_{\nu_2]},
\label{eq:Ramp}
\end{equation}
where $\mu_1\mu_2$ are associated with the left-movers, while $\nu_1\nu_2$ are associated
with the right-movers.  Here the tilde indicates that the curvature tensor is given in the Einstein frame, which is the natural frame corresponding to the string amplitudes as the two-point functions are diagonal between $h_{\mu\nu}$, $b_{\mu\nu}$ and $\phi$.  Noting the polarization decomposition (\ref{eq:thetapol}), we see that $\bar{\tilde R}$ can be written as
\begin{equation}
\bar{\tilde R}_{\mu_1\mu_2}{}^{\nu_1\nu_2}=\tilde R_{\mu_1\mu_2}{}^{\nu_1\nu_2}
+e^{-\phi/2}\nabla_{[\mu_1}H_{\mu_2]}{}^{\nu_1\nu_2}
-\delta_{[\mu_1}{}^{[\nu_1}\nabla_{\mu_2]}\nabla^{\nu_2]}\phi,
\label{eq:Rbar}
\end{equation}
which is the linearised form of a connection with torsion.  Note that we have introduced the $e^{-\phi/2}$
factor in front of $\nabla H$, which has no effect on the four-point function, but ensures that $H$ has the proper weight in the Einstein frame.  In addition, the Bianchi identity $dH=0$ ensures that
\begin{equation}
\bar{\tilde R}_{\mu_1\mu_2\nu_1\nu_2}(H)=\bar{\tilde R}_{\nu_1\nu_2\mu_1\mu_2}(-H),
\label{eq:Rflip}
\end{equation}
which is compatible with worldsheet parity.

Although the string amplitudes directly reproduce the effective action in the Einstein frame, we can transform into the string frame by taking $\tilde g_{\mu\nu}=e^{\phi/2}g_{\mu\nu}$.  In the string frame, the curvature tensor with torsion, (\ref{eq:Rbar}), takes a particularly simple form as the curvature of a connection with torsion $\Omega_+=\omega+\ft12\mathcal H$, where $\mathcal H$ is viewed as a one-form taking values in the tangent space
\begin{equation}
\mathcal H^{\alpha\beta}=H_\mu{}^{\alpha\beta}dx^\mu.
\end{equation}
The curvature computed out of $\Omega_+$ is then
\begin{equation}
R(\Omega_+)=R+\ft12d\mathcal H+\ft14\mathcal H\wedge\mathcal H,
\label{eq:Romegaplus}
\end{equation}
which has the component form
\begin{equation}
R(\Omega_+)_{\mu\nu}{}^{\alpha\beta}=R_{\mu\nu}{}^{\alpha\beta}
+\nabla_{[\mu}H_{\nu]}{}^{\alpha\beta}+\ft12H_{[\mu}{}^{\alpha\gamma}H_{\nu]\gamma}{}^\beta.
\end{equation}
In the string frame, the tree-level effective action reproducing the eight-derivative four-point function then takes the form  \cite{Gross:1986iv,Gross:1986mw}
\begin{equation}
\mathcal L_{\rm tree}=\sqrt{-g}\,e^{-2\phi}\left[R+4\partial\phi^2-\fft1{12}H^2+\fft{\zeta(3)}{3\cdot2^{11}}\alpha'^3
(t_8t_8-\ft14\epsilon_{8}\epsilon_{8}) R(\Omega_+)^4+\cdots\right],
\end{equation}
where we have restored the tree-level numerical factor.  Here we should explain our notation for the $\epsilon_n\epsilon_n$ tensor.  While $\epsilon_{10}$ is the fully antisymmetric tensor in ten dimensions, a repeated $\epsilon_n\epsilon_n$ will denote the antisymmetric delta function with $n$ pairs of indices
\begin{equation}
\epsilon_{n\,\mu_1\cdots\mu_n}\epsilon_n^{\nu_1\cdots\nu_n}=n!\delta_{\mu_1}^{[\nu_1}\cdots\delta_{\mu_n}^{\nu_n]}=-\fft1{m!}\epsilon_{\alpha_1\cdots\alpha_m\mu_1\cdots\mu_n}\epsilon^{\alpha_1\cdots\alpha_m\nu_1\cdots\nu_n}\qquad \mbox{with}\quad m+n=10.
\label{eq:epsepsdef}
\end{equation}
The sign arises because we are working with a Lorentzian signature in ten dimensions.  

At this point, several comments are in order.  Firstly, we have included the $\epsilon_{8}\epsilon_{8}R(\Omega_+)^4$ term, even though it does not contribute to the four-point function.  This term is implicit in the Green-Schwarz formalism and moreover is needed for agreement with the $\sigma$-model approach
\cite{Grisaru:1986vi,Freeman:1986zh}.  Secondly, while the four-point function is only sensitive to the linearised curvature, the natural object in the effective action is the full non-linear curvature tensor $R(\Omega_+)$.  This non-linear completion was anticipated in \cite{Kehagias:1997cq} and further support for its structure was given in \cite{Liu:2013dna}.  Both of these features will be seen directly at the level of the five-point function.  Finally, although we have focused on the NSNS sector, the full tree-level quartic effective action has been computed using pure-spinor methods \cite{Policastro:2006vt,Policastro:2008hg}.

Just as for the open string, the closed-string tree and one-loop four-point functions are based on identical kinematical factors.  In particular, we have \cite{Green:1999pv}
\begin{equation}
M_4^{\mathrm{loop}}\sim\alpha'^3\left(\fft{2\pi^2}{3}+\alpha'^3\fft{\zeta(2)\zeta(3)}2s_{12}s_{23}s_{13}+\cdots\right)\left|s_{12}s_{23}A_{\mathrm{YM}}(1,2,3,4)\right|^2.
\label{eq:M4l}
\end{equation}
This time, the eight-derivative term is leading, corresponding to the one-loop effective action
\begin{equation}
\mathcal L_{\mathrm{loop}}=\sqrt{-g}\left[\fft{\pi^2}{9\cdot2^{11}}\alpha'^3
(t_8t_8\pm\ft14\epsilon_{8}\epsilon_{8}) R(\Omega_+)^4+\cdots\right],
\label{eq:Lloop}
\end{equation}
where the top sign corresponds to the IIA string.  The difference in sign between the IIA and IIB strings arises because of the difference in GSO projections (or equivalently the difference in $SO(8)$ chiralities in the Green-Schwarz formalism).  Again, the $\epsilon_{10}\epsilon_{10}$ term is not visible at the level of the four-point function, but can be seen in the five-point function.

\subsection{The one-loop five-point function}

As indicated above, many of the additional features of the eight-derivative couplings are not visible at the level of the four-point function.  Thus we now consider the additional input arising from five-point functions.  The tree-level open string five-point amplitude was computed in \cite{Kitazawa:1987xj,Medina:2002nk,Barreiro:2005hv,Stieberger:2006te} using the covariant approach, and the one-loop amplitude was computed in \cite{Frampton:1985uw,Frampton:1986ea,Frampton:1986gi,Lam:1986kg} in the Green-Schwarz formalism and in \cite{Montag:1992dm} in the even-even sector in the covariant formalism.  More recently, following the development of the pure-spinor formalism \cite{Berkovits:2000fe}, enormous progress has been made in computing string amplitudes \cite{Mafra:2009wq}, including higher-point trees  \cite{Mafra:2011nv,Mafra:2011nw} and loops \cite{Mafra:2012kh,Mafra:2018nla,Mafra:2018pll,Mafra:2018qqe}.  More concretely, the five-point open-string tree amplitude was revisited in \cite{Mafra:2009bz}, and the closed-string tree amplitude follows from the KLT relations.  The five-point open-string loop amplitude was reexamined in \cite{Mafra:2009wi}, and in principle the closed-string loop amplitude follows by using chiral splitting and exploiting the double-copy.  Alternatively, the closed-string five-point loop amplitude can be directly computed \cite{Richards:2008jg}.

Of course, the string amplitudes themselves are not the end of the story, as we are interested in recreating the eight-derivative effective-action from the amplitudes.  The one-loop effective action was investigated in \cite{Richards:2008jg} for the scattering of five gravitons and in \cite{Peeters:2001ub,Richards:2008sa} for anti-symmetric tensors and gravitons.  The starting point is the closed-string one-loop five-point amplitude, given here for the CP-even sector and at the eight-derivative level \cite{Richards:2008jg}
\begin{align}
M_5^{\mathrm{loop}}&\sim-\fft1{s_{12}}\Bigl[k_1\cdot e_2 t_8(e_1,k_1+k_2,e_3,k_3,e_4,k_4,e_5,k_5)-k_2\cdot e_1 t_8(e_2,k_1+k_2,e_3,k_3,e_4,k_4,e_5,k_5)\nn\\
&\kern4em-e_1\cdot e_2 t_8(k_1,k_2,e_3,k_3,e_4,k_4,e_5,k_5)-s_{12}t_8(e_1,e_2,e_3,k_3,e_4,k_4,e_5,k_5)\Bigr]\nn\\
&\qquad\times\Bigl[k_1\cdot\bar e_2 t_8(\bar e_1,k_1+k_2,\bar e_3,k_3,\bar e_4,k_4,\bar e_5,k_5)-k_2\cdot\bar e_1 t_8(\bar e_2,k_1+k_2,\bar e_3,k_3,\bar e_4,k_4,\bar e_5,k_5)\nn\\
&\kern4em-\bar e_1\cdot\bar e_2 t_8(k_1,k_2,\bar e_3,k_3,\bar e_4,k_4,\bar e_5,k_5)-s_{12}t_8(\bar e_1,\bar e_2,\bar e_3,k_3,\bar e_4,k_4,\bar e_5,k_5)\Bigr]\nn\\
&\qquad-e_1\cdot\bar e_2 t_8(e_2,k_2,e_3,k_3,e_4,k_4,e_5,k_5)t_8(\bar e_1,k_1,\bar e_3,k_3,\bar e_4,k_4,\bar e_5,k_5)\nn\\
&\qquad-e_2\cdot\bar e_1 t_8(e_1,k_1,e_3,k_3,e_4,k_4,e_5,k_5)t_8(\bar e_2,k_2,\bar e_3,k_3,\bar e_4,k_4,\bar e_5,k_5)\nn\\
&\quad\mp\fft14\Bigl[s_{12}\epsilon_8(e_1,e_2,e_3,k_3,e_4,k_4,e_5,k_5)\epsilon_8(\bar e_1,\bar e_2,\bar e_3,k_3,\bar e_4,k_4,\bar e_5,k_5)\nn\\
&\kern4em+e_1\cdot\bar e_2\epsilon_8(e_2,k_2,e_3,k_3,e_4,k_4,e_5,k_5)\epsilon_8(\bar e_1,k_1,\bar e_3,k_3,\bar e_4,k_4,\bar e_5,k_5)\nn\\
&\kern4em+e_2\cdot\bar e_1\epsilon_8(e_1,k_1,e_3,k_3,e_4,k_4,e_5,k_5)\epsilon_8(\bar e_2,k_2,\bar e_3,k_3,\bar e_4,k_4,\bar e_5,k_5)\Bigr]\nn\\
&\quad+\mbox{9 more in the other $s_{ij}$ channels}\nn\\
&\quad-e_1\cdot\bar e_1 t_8(e_2,k_2,e_3,k_3,e_4,k_4,e_5,k_5)t_8(\bar e_2,k_2,\bar e_3,k_3,\bar e_4,k_4,\bar e_5,k_5)\nn\\
&\quad\mp\fft14e_1\cdot\bar e_1\epsilon_8(e_2,k_2,e_3,k_3,e_4,k_4,e_5,k_5)\epsilon_8(\bar e_2,k_2,\bar e_3,k_3,\bar e_4,k_4,\bar e_5,k_5)\nn\\
&\quad+\mbox{4 more for vertices $2,\ldots,5$}.
\label{eq:M5l}
\end{align}
Here the top signs correspond to the IIA string while the bottom signs correspond to the IIB string.  This amplitude was computed in \cite{Richards:2008jg} in the Green-Schwarz formalism, and we have explicitly written out the combination $t_8\pm\fft12\epsilon_8$.  Here $\epsilon_8$ is the fully antisymmetric tensor in eight dimensions with Euclidean signature.  Note that the final three lines in (\ref{eq:M5l}) correspond to trace polarizations, which are important when considering amplitudes involving dilatons.  In particular, they were not included in \cite{Richards:2008jg,Richards:2008sa}, which only dealt with graviton and anti-symmetric tensor amplitudes.  The normalization here is such that the right-hand side of (\ref{eq:M5l}) reproduces
\begin{equation}
\mathcal L\sim\fft2{2^4\cdot 4!}(t_8t_8\pm\ft14\epsilon_{8}\epsilon_{8})R^4+\cdots,
\end{equation}
where $1/4!$ is a symmetry factor, and each factor of two in the denominator arises from the normalization of the linearised Riemann tensor, $R_{abcd}=2k_{[a}e_{b]}k_{[c}\bar e_{d]}$.

Note that the $\epsilon_8\epsilon_8$ terms in (\ref{eq:M5l}) arise directly from the odd-odd spin structure sector in the covariant worldsheet approach.  This amplitude first appears at the level of the five-point function, as ten fermion zero modes need to be soaked up on each side of the string.  The odd-odd amplitude takes the simple form \cite{Peeters:2001ub,Liu:2013dna}
\begin{equation}
M_5^{\textrm{o-o}}\sim\mp\fft14\epsilon_9(e_1,e_2,k_2,e_3,k_3,e_4,k_4,e_5,k_5)\epsilon_9(\bar e_1,\bar e_2,k_2,\bar e_3,k_3,\bar e_4,k_4,\bar e_5,k_5),
\label{eq:M1oo}
\end{equation}
in the same normalization as (\ref{eq:M5l}).  Here $\epsilon_9$ is given in Euclidean signature, and $\epsilon_9\epsilon_9$ arises from a single bosonic zero mode contraction between two $\epsilon_{10}$ tensors.  This amplitude is particularly simple because of the absence of any odd-odd four-point amplitude that would give rise to pole terms from intermediate state particle exchange.  Consistency between the Green-Schwarz and covariant amplitudes demand that the summed $\epsilon_8\epsilon_8$ terms in (\ref{eq:M5l}) match the single $\epsilon_9\epsilon_9$ expression in (\ref{eq:M1oo}), and this can be shown explicitly using on-shell five-point kinematics and the schematic decomposition $\epsilon_9\epsilon_9\sim\sum\delta_{ij}\epsilon_8\epsilon_8$.

As shown in \cite{Richards:2008jg}, the five-graviton scattering amplitude reproduces the expected one-loop $R^4$ terms (\ref{eq:Lloop}) at the non-linear level, while the amplitudes involving anti-symmetric tensors and gravitons give rise to new terms in the odd-odd spin structure sector \cite{Richards:2008sa,Liu:2013dna}
\begin{align}
\mathcal L_{\mathrm{loop}}&=\sqrt{-g}\biggl[\fft{\pi^2}{9\cdot2^{11}}\alpha'^3
\left(t_8t_8R(\Omega_+)^4\pm\ft14\epsilon_{8}\epsilon_{8} R(\Omega_+)^4\pm\ft13\epsilon_{9}\epsilon_{9}H^2R(\Omega_+)^3\mp\ft49\epsilon_{9}\epsilon_{9}H^2(\nabla H)^2R(\Omega_+)+\cdots\right)\nn\\
&\kern4em+\cdots\biggr]-\fft{(2\pi)^2}{3\cdot2^6}\alpha'^3B_2\wedge\Bigl[\tr R(\Omega_+)^4-\ft14(\tr R(\Omega_+)^2)^2\Bigr]_{\mbox{\scriptsize even(odd) in $B_2$ for IIA(IIB)}},
\label{eq:loopea}
\end{align}
where
\beq
\epsilon_{9}\epsilon_{9}H^2R(\Omega_+)^3 \equiv-\epsilon_{\alpha\mu_0\mu_1\cdots\mu_8}\epsilon^{\alpha\nu_0\nu_1\cdots\nu_8}
H^{\mu_1\mu_2}{}_{\nu_0}H_{\nu_1\nu_2}{}^{\mu_0}
R^{\mu_3\mu_4}{}_{\nu_3\nu_4}(\Omega_+)
R^{\mu_5\mu_6}{}_{\nu_5\nu_6}(\Omega_+)R^{\mu_7\mu_8}{}_{\nu_7\nu_8}(\Omega_+),
\label{eq:eeH2R3}
\eeq
and
\begin{equation}
\epsilon_{9}\epsilon_{9}H^2(\nabla H)^2R(\Omega_+)\equiv-\epsilon_{\alpha\mu_0\mu_1\cdots\mu_8}\epsilon^{\alpha\nu_0\nu_1\cdots\nu_8}
H^{\mu_1\mu_2\mu_0}H_{\nu_1\nu_2\nu_0}\nabla^{\mu_3}H^{\mu_4}{}_{\nu_3\nu_4}
\nabla^{\mu_5}H^{\mu_6}{}_{\nu_5\nu_6}R^{\mu_7\mu_8}{}_{\nu_7\nu_8}(\Omega_+).
\label{eq:eeH2NH2R}
\end{equation}
Note that, compared to \cite{Liu:2013dna}, we have simplified the $\epsilon\epsilon H^2(\nabla H)^2R$ term using the identity
\begin{align}
&\epsilon_{\alpha\mu_0\mu_1\cdots\mu_8}\epsilon^{\alpha\nu_0\nu_1\cdots\nu_8}
H^{\mu_1\mu_2}{}_{\nu_0}H_{\nu_1\nu_2}{}^{\mu_0}
\nabla^{\mu_3}H^{\mu_4}{}_{\nu_3\nu_4}
\nabla^{\mu_5}H^{\mu_6}{}_{\nu_5\nu_6}R^{\mu_7\mu_8}{}_{\nu_7\nu_8}\nn\\
&\kern9em=\ft79
\epsilon_{\alpha\mu_0\mu_1\cdots\mu_8}\epsilon^{\alpha\nu_0\nu_1\cdots\nu_8}
H^{\mu_1\mu_2\mu_0}H_{\nu_1\nu_2\nu_0}\nabla^{\mu_3}H^{\mu_4}{}_{\nu_3\nu_4}
\nabla^{\mu_5}H^{\mu_6}{}_{\nu_5\nu_6}R^{\mu_7\mu_8}{}_{\nu_7\nu_8}+\cdots,
\end{align}
which is valid at the level of the on-shell five-point function.  While this effective Lagrangian is only complete up to the five-point function, there is evidence that the CP-odd sector as well as the even-even spin structure term is complete at it stands \cite{Liu:2013dna}.  On the other hand, the odd-odd spin structure term will receive contributions from six and presumably higher-point functions, as can be seen from the reduction to six dimensions.

\subsection{The tree-level five-point function}
\label{sec:tree5}

We now return to our main focus, which is on the tree-level five-point function and its implication on the eight-derivative effective action.  The closed-string tree-level amplitude can be obtained using the KLT relations \cite{Kawai:1985xq} and, at the eight-derivative level, takes the form \cite{Schlotterer:2012ny,Boels:2013jua,Green:2013bza}
\begin{equation}
M_5^{\mathrm{tree}}\sim\alpha'^32\zeta(3)\begin{pmatrix}\bar A_{\mathrm{YM}}(1,2,3,5,4)\\\bar A_{\mathrm{YM}}(1,3,2,5,4)\end{pmatrix}^TS_0M_3\begin{pmatrix}A_{\mathrm{YM}}(1,2,3,4,5)\\A_{\mathrm{YM}}(1,3,2,4,5)\end{pmatrix}.
\label{eq:M5t}
\end{equation}
Here $A_{\mathrm{YM}}(1,2,3,4,5)$ is the Yang-Mills five-point amplitude and $S_0$ and $M_3$ are $2\times2$ matrices
\begin{equation}
S_0=\begin{pmatrix}s_{12}(s_{13}+s_{23})&s_{12}s_{13}\\s_{12}s_{13}&s_{13}(s_{12}+s_{23})\end{pmatrix},\qquad
M_3=\begin{pmatrix}m_{11}&m_{12}\\m_{21}&m_{22}\end{pmatrix},
\end{equation}
with
\begin{align}
m_{11}&\kern5em=s_3\left(-s_1(s_1+2s_2+s_3)+s_3s_4+s_4^2\right)+s_1s_5(s_1+s_5),\nn\\
m_{12}&\kern5em=-s_{13}s_{24}(s_1+s_2+s_3+s_4+s_5),\nn\\
m_{21}&=m_{12}\big|_{2\leftrightarrow3}=-s_1s_3(s_{13}+s_2+s_{24}+s_4+s_5),\nn\\
m_{22}&=m_{11}\big|_{2\leftrightarrow3}=s_{24}\left(-s_{13}(s_{13}+2s_2+s_{24})+s_{24}s_4+s_4^2\right)+s_{13}s_5(s_{13}+s_5),
\end{align}
where $s_{ij}=k_i\cdot k_j$ and $s_i\equiv s_{i,i+1}$.  In general, we can form ten Mandelstam invariants $s_{ij}$ with $i<j$.  However, massless five-particle kinematics allows us to reduce this to five independent variables, say $s_1,\ldots,s_5$.  The other five variables can then be expressed as
\begin{align}
s_{13}&=s_4-s_1-s_2,&s_{14}&=s_2-s_4-s_5,&s_{24}&=s_5-s_2-s_3,\nn\\
s_{25}&=s_3-s_1-s_5,&s_{35}&=s_1-s_3-s_4.
\end{align}
The explicit form of $A_{\mathrm{YM}}(1,2,3,4,5)$ is available from \cite{PSS-website}, and contains poles corresponding to factorization on intermediate gluon states.  The KLT product (\ref{eq:M5t}) superficially appears to have pole-squared terms, but explicit evaluation with Mathematica \cite{Mathematica} shows that no such terms are present.  Single pole terms, however, do remain, corresponding to factorization on intermediate NSNS closed string states.

From an effective action point of view, the five-point amplitude receives two types of contributions.  The first type corresponds to intermediate closed-string exchange in the $s_{ij}$ channel connecting a leading-order three-point amplitude for particles $i$, $j$ and the intermediate state to a tree-level $\alpha'^3$ four-point amplitude for the intermediate state with the remaining three external particles.  Here the intermediate closed-string state could be any one of the massless NSNS states, namely the graviton, anti-symmetric tensor or dilaton.  The second type of contribution is the actual five-point contact interaction, and this is what we are mostly interested in since it corresponds to additional couplings beyond what appears in the quartic effective action.

For the one-loop amplitude, the subtraction of pole terms from the intermediate channel particle exchanges was carried out in \cite{Richards:2008jg} for the five-graviton amplitude and in \cite{Richards:2008sa} for the mixed anti-symmetric tensor and graviton amplitudes.  This process is similar for all combination of external particles, and the resulting contact interaction is then used to recreate the one-loop effective action (\ref{eq:loopea}).  In principle, we can repeat this subtraction procedure for the tree-level amplitude.  However, we find it convenient to take a shortcut of working with the difference between tree and loop amplitudes
\begin{equation}
\Delta M_5=\left.M_5^{\mathrm{tree}}-M_5^{\mbox{\scriptsize IIB loop}}\right|_{\mathcal O(\alpha'^3)}.
\label{eq:DM5}
\end{equation}
Note that only the $\mathcal O(\alpha'^3)$ (ie eight-derivative) contributions are to be included in this difference.  Moreover, the tree and loop factors, proportional to $2\zeta(3)$ and $2\pi^2/3$, respectively, have been removed so the difference would vanish for identical kinematics.  Although the tree and loop amplitudes differ by a dilaton factor $e^{2\phi}$, the amplitudes correspond to a perturbative expansion with vanishing dilaton vev so their difference is still meaningful.  Furthermore, while the tree-level amplitude is identical for the IIA and IIB strings, we compare with the one-loop IIB amplitude because of the expected $S$-duality invariance of the IIB string.  In particular, at the level of the four-point function, the difference $\Delta M_4$ vanishes identically for any combination of external NSNS states, as the $\mathcal O(\alpha'^3)$ kinematical factors in (\ref{eq:M4t}) and (\ref{eq:M4l}) are fixed by supersymmetry and hence are forced to be identical.  When combined with non-perturbative contributions, this leads to the $SL(2,\mathbb Z)$ invariant combination \cite{Green:1997tv, Green:1997di}
\begin{equation}
\mathcal L_{R^4}\sim\alpha'^3\mathcal E_{3/2}(\tau,\bar\tau)(t_8t_8-\ft14\epsilon_8\epsilon_8)R^4,
\label{eq:SL2}
\end{equation}
at least in the purely gravitational sector.

Returning to the five-point function, (\ref{eq:DM5}), this difference turns out to be non-vanishing in general when the external closed-string polarizations are unspecified.  However, all pole terms cancel so $\Delta M_5$ is purely a contact interaction.  This greatly simplifies our task of obtaining the tree-level effective action as we will not have to subtract out any underlying pole terms.  Even without any poles, the expression for $\Delta M_5$ in terms of polarizations and momenta is quite long, so we have used Mathematica \cite{Mathematica} for simplification.

Although $\Delta M_5$ is not identically zero, it does vanish in the case of five graviton scattering where all five external polarizations are taken to be symmetric and trace-free.  (Note that we are not considering any dilaton external states where the difference between tree and loop amplitudes will be manifested.)  As a result, the $SL(2,\mathbb Z)$ invariant structure (\ref{eq:SL2}) continues to hold at the next non-linear level, as expected.

The more interesting case to consider is the scattering of two anti-symmetric tensors and three gravitons.  Here it turns out that $\Delta M_5$ does not vanish, and moreover has a long and unilluminating expansion in terms of external polarizations and momenta.  Nevertheless, we expect that this amplitude can be reproduced by a suitable quintic effective action made out of gauge-invariant combinations of $H^2R^3$.  We thus start by constructing a complete basis of scalar $H^2R^3$ invariants and then computing the five-point amplitudes corresponding to these invariants.  After this, we finally decompose $\Delta M_5$ into a linear combination of these invariants and thereby deduce the effective action.

In order to obtain a complete basis of $H^2R^3$ invariants, we first consider the tensor decomposition of $H^2$ and $R^3$ and then look for singlet combinations under the Lorentz group $SO(1,9)$.  We start with the symmetric tensor product of $H^2$
\begin{align}
\otimes^2H&\to[0,0,0,0,0]+[2,0,0,0,0]+[0,2,0,0,0]+
[0,0,0,1,1] +[0,0,2,0,0]+[1,0,0,0,2]+[1,0,0,2,0]\nn\\
&\kern3.2emH^2\kern1.8em+\kern1.2em H^2_{(\mu\nu)}\kern1.2em+\kern.7em H^2_{[\mu\nu][\rho\sigma]}\kern.6em+\kern.8em H^2_{[\mu\nu\rho\sigma]}\kern.8em+\kern.1em H^2_{[\mu\nu\lambda][\alpha\beta\gamma]}\kern.1em+\kern3em H^2_{[\mu\nu\lambda\rho\sigma]\zeta}.
\end{align}
The representations are given in terms of Dynkin labels and also shown schematically on the second line.  Note that the last two irreducible representations correspond to the self-dual and anti-self-dual components of the five-form indices.  This decomposition now singles out the corresponding representations in the decomposition of $R^3$ that can be used to form overall singlets.  The symmetric tensor product of $R^3$ is more involved, but can be obtained with the assistance of the LiE computer algebra package \cite{LiE1,LiE2,LiE3}.  Since we are free to ignore Ricci terms at this order, we actually consider the symmetric product of three $[0,2,0,0,0]$ irreducible representations, corresponding to the Weyl tensor.  The relevant terms for forming singlet combinations with $H^2$ are shown in Table~\ref{tbl:R3dec}, where the notation parallels that of \cite{Coimbra:2017fqj} with the exception of $Q^i$, which did not appear there.

\begin{table}[t]
\begin{center}
\begin{tabular}{ccc}
Projection of $\otimes^3R$&Representation&Multiplicity\\
\hline
$S^i$&$[0,0,0,0,0]$&2\\
$W^i$&$[2,0,0,0,0]$&3\\
$X^i$&$[0,2,0,0,0]$&8\\
$T^i$&$[0,0,0,1,1]$&3\\
$Q^i$&$[0,0,2,0,0]$&6\\
$V_+^i$&$[1,0,0,0,2]$&2\\
$V_-^i$&$[1,0,0,2,0]$&2
\end{tabular}
\caption{Irreducible $SO(1,9)$ representations and their multiplicities in the decomposition of $\otimes^3R$ that can be used to form singlets with $H^2$.}
\label{tbl:R3dec}
\end{center}
\end{table}

The main result from Table~\ref{tbl:R3dec} is that there are two scalar $R^3$ invariants $S^i$, three two-index invariants $W^i$, eleven four-index invariants $\{X^i,T^i\}$ and eight parity conserving six-index invariants $\{Q^i,V^i\}$ where $V^i=V_+^i+V_-^i$.  This requires us to introduce a total of 24 basis terms to span the full set of possible $H^2R^3$ combinations in the CP-even sector.  Although the above decomposition is given in a basis of irreducible representations, we find the use of a reducible basis to be more straightforward as this obviates the need to project out traces and mixed tensor structures.  We then take
\begin{equation}
\mathcal L_{\mathrm{basis}}=\sqrt{-g}\left[a_iH^2\tilde S^i+b_iH^\mu{}_{ab}H^{\nu\,ab}\tilde W_{\mu\nu}^i+c_iH^{\mu\nu}{}_aH^{\rho\sigma\,a}\tilde X^i_{\mu\nu\rho\sigma}+d_iH^{\mu\nu\lambda}H^{\rho\sigma\zeta}\tilde Q^i_{\mu\nu\lambda\rho\sigma\zeta}\right]
\end{equation}
where the specific $R^3$ combinations are given in Appendix~\ref{app:R3}.

It is now a straightforward exercise to take the difference in amplitudes, (\ref{eq:DM5}), and decompose it into the above basis.  The result is given in the first line of Table~\ref{tbl:bde}.  Although this decomposition is not particularly illuminating by itself, it can be partially rewritten using the $t_8$ and $\epsilon_{10}$ tensors, as shown in the additional lines of the table.  Here the $\epsilon_{9}\epsilon_{9}H^2R^3$ term has the same form as (\ref{eq:eeH2R3}), while the $t_8t_8H^2R^3$ term has the explicit index structure
\begin{equation}
t_8t_8H^2R^3\equiv t_{8\,\mu_1\cdots\mu_8}t_8^{\nu_1\cdots\nu_8}H^{\mu_1\mu_2\alpha}H_{\nu_1\nu_2\alpha}R^{\mu_3\mu_4}{}_{\nu_3\nu_4}R^{\mu_5\mu_6}{}_{\nu_5\nu_6}R^{\mu_7\mu_8}{}_{\nu_7\nu_8}.
\end{equation}
This term was initially proposed in \cite{Grimm:2017okk} in order to recover four-dimensional supersymmetry based on Calabi-Yau compactification.  As indicated in Table~\ref{tbl:bde}, the difference $\Delta M_5$ can be written in terms of $t_8t_8H^2R^3$ and $\epsilon_9\epsilon_{9}H^2R^3$ and a set of terms of the form $H^{\mu\nu\lambda}H^{\rho\sigma\zeta}\tilde Q^i_{\mu\nu\lambda\rho\sigma\zeta}$ where $H^2$ is fully uncontracted.  Adding back in the IIB one-loop amplitude and restoring the appropriate tree-level factor of $2\zeta(3)$ then gives
\begin{align}
\mathcal L_{\mathrm{tree}}&=\sqrt{-g}\,e^{-2\phi}\biggl[\fft{\zeta(3)}{3\cdot2^{11}}\alpha'^3
\Bigl(t_8t_8R(\Omega_+)^4-\ft14\epsilon_{8}\epsilon_{8} R(\Omega_+)^4-2t_8t_8H^2R(\Omega_+)^3-\ft16\epsilon_{9}\epsilon_{9}H^2R(\Omega_+)^3\nn\\
&\kern10em+8\cdot4!\sum_id_iH^{\mu\nu\lambda}H^{\rho\sigma\zeta}\tilde Q^i_{\mu\nu\lambda\rho\sigma\zeta}+\cdots\Bigr)+\cdots\biggr],
\label{eq:treeea}
\end{align}
where
\begin{equation}
\label{eq:di}
\{d_i\}=(1,-\ft14,0,\ft13,1,\ft14,-2,\ft18),
\end{equation}
and the $R^3$ combinations $\tilde Q^i_{\mu\nu\lambda\rho\sigma\zeta}$ can be found in Appendix~\ref{app:R3}.  Note that we have chosen to write the $H^2R^3$ terms using the curvature $R(\Omega_+)$ given in (\ref{eq:Romegaplus}), as the distinction between $H^2R^3$ and $H^2R(\Omega_+)^3$ only arises at the level of the six-point function and beyond.

\begin{table}[t]
\begin{center}
\setlength\tabcolsep{2pt}
\renewcommand\arraystretch{1.4}
\begin{tabular}{l|cc|ccc|ccccccccccc|cccccccc}
&$a_1$&$a_2$&$b_1$&$b_2$&$b_3$&$c_1$&$c_2$&$c_3$&$c_4$&$c_5$&$c_6$&$c_7$&$c_8$&$c_9$&$c_{10}$&$c_{11}$&$d_1$&$d_2$&$d_3$&$d_4$&$d_5$&$d_6$&$d_7$&$d_8$\\
\hline
$\Delta M_5$&$\fft1{72}$&$\fft1{36}$&$\fft14$&$-\fft14$&$\fft12$&0&0&0&0&0&0&0&$1$&$-2$&$\fft12$&$-\fft12$&$1$&$-\fft12$&$-\fft12$&0&0&0&0&0\\
\hline
$-\fft1{4\cdot4!}t_8t_8H^2R^3$&0&0&0&0&0&$-\fft1{32}$&$\fft12$&$-\fft1{16}$&0&$-\fft12$&$-\fft12$&$\fft12$&$1$&$-1$&$\fft14$&0&0&0&0&0&0&0&0&0\\
$\fft1{48\cdot4!}\epsilon_{9}\epsilon_{9}H^2R^3$&$\fft1{72}$&$\fft1{36}$&$\fft14$&$-\fft14$&$\fft12$&$\fft1{32}$&$-\fft12$&$\fft1{16}$&0&$\fft12$&$\fft12$&$-\fft12$&0&$-1$&$\fft14$&$-\fft12$&0&$-\fft14$&$-\fft12$&$-\fft13$&$-1$&$-\fft14$&$2$&$-\fft18$
\end{tabular}
\caption{The difference $\Delta M_5$ between the tree and loop amplitudes decomposed into the 24 basis elements.  The decomposition of relevant $H^2R^3$ terms is also shown for comparison.}
\label{tbl:bde}
\end{center}
\end{table}

While the tree-level effective Lagrangian, (\ref{eq:treeea}), fully captures the $H^2R^3$ terms, it is important to keep in mind that it is still incomplete at the quintic level, as we have not considered $H^2(\nabla H)^2R$ terms nor have we considered couplings involving the dilaton.  Since the tree-level five-point amplitude, (\ref{eq:M5t}), includes all polarizations in the NSNS sector, it would be straightforward to extend (\ref{eq:treeea}) to the entire NSNS sector at the quintic level.  However, such additional couplings will not be important for the reductions considered below.

This explicit construction of the tree-level $H^2R^3$ couplings confirms the prediction of \cite{Grimm:2017okk} that a tree-level term of the form $\delta\mathcal J=-2t_8t_8H^2R^3$ is needed for the reduction of IIA theory on $CY_3$ to be supersymmetric in four dimensions.  In addition, it demonstrates that the $\epsilon\epsilon H^2R^3$ coupling, which arises from the odd-odd sector in a covariant one-loop computation \cite{Liu:2013dna}, is also present at tree-level, however with a relative factor of one half compared to the one-loop coupling.  Because of this relative factor, and because of the additional terms in (\ref{eq:treeea}), we see that $SL(2,\mathbb Z)$ invariance of the IIB theory cannot be obtained simply by the use of a single automorphic function of the form (\ref{eq:SL2}) once additional fields beyond the graviton are included.  We will return to this point below, when we consider $S$-duality of IIB theory with higher derivative terms.

Finally, the presence of the fully uncontracted $H^2$ terms in the second line of (\ref{eq:treeea}) came as somewhat of a surprise since they are not present in such a form at the one-loop level.  Actually, the expansion of $\epsilon_{9}\epsilon_{9}H^2R^3$ contains such terms, as can be seen in the last row of Table~\ref{tbl:bde}.  However, this expansion yields a different combination than that of the left over terms in the tree-level effective action.  Along these lines, we have attempted but failed to rewrite the fully uncontracted $H^2$ terms in terms of invariants built out of $t_8$, $\epsilon_{10}$ and the metric tensor.  Of course, we have not been exhaustive in doing so, and it remains an open question whether any further simplification of (\ref{eq:treeea}) is possible.

\section{Testing the couplings: K3 reduction}
\label{sec:K3}

Starting from the tree-level closed string five-point amplitude in the NSNS sector, we have constructed a supergravity effective action, (\ref{eq:treeea}), that reproduces the tree-level eight derivative couplings up to $H^2R^3$ order.  Although this is not the full quintic effective action, as it lacks $H^2(\nabla H)^2R$ terms as well as potentially dilaton terms, it nevertheless indicates the presence of additional couplings that can be investigated upon dimensional reduction.  For example, reduction on K3 gives rise to four-derivative couplings (among others) arising from factorized $(\mbox{four derivative})\times R^2$ terms in ten dimensions and reduction on a Calabi-Yau threefold gives rise to renormalized two-derivative interactions arising from factorized $(\mbox{two derivative})\times R^3$ terms.  Supersymmetry provides strong constraints on these sort of terms, and this provides a convenient check on the ten-dimensional couplings.

Here we consider the K3 reduction of the quintic couplings and consider both tree and loop terms for completeness.  In order to reduce an eight-derivative term in ten dimensions to a four-derivative term, we must soak up four derivatives using the curvature of K3.  This means we are only sensitive to terms involving a factorized $R_{\mu\nu\rho\sigma}^2$.  Gauge invariant four-derivative contact terms built from the antisymmetric tensor and metric range schematically from $R^2$ up to $H^4$, and hence can be probed completely using four-point functions in six dimensions.  When combined with $R_{\mu\nu\rho\sigma}^2$, this lifts to six-point functions in ten dimensions, which we have not explored.  In particular, the quintic effective action, (\ref{eq:treeea}), can be used to constrain no more than three-point contact terms in six dimensions, so we will be unable to provide a full test of the couplings.  Nevertheless, the partial information is still illuminating.

\subsection{The six-dimensional tree-level couplings}

We begin with an examination of the tree-level couplings.  While four-derivative couplings are generally non-vanishing in theories preserving 16 real supercharges, the tree-level interactions within the six-dimensional gravity multiplet begin at the four-point, eight-derivative level, just as in the ten-dimensional type-II case.  One quick way to see this is to consider the orbifold limit of K3.  The tree-level amplitude in the untwisted sector, which is where the gravity multiplet lives, is then identical to that on $T^4$, and it is well known that the latter only receives corrections starting at the eight-derivative level.

On the other hand, the reduction of the tree-level quintic action, (\ref{eq:treeea}), on K3 yields
\begin{equation}
\mathcal L_{\mathrm{tree}}^{d=6}=\sqrt{-g}e^{-2\phi}\left[R+4\partial\phi^2-\ft1{12}H^2\right]+\mathcal L_{\mathrm{tree}}^{\partial^4},
\end{equation}
where
\begin{equation}
\mathcal L_{\mathrm{tree}}^{\partial^4}=\sqrt{-g}e^{-2\phi}\alpha\Bigl((t_4t_4-\ft14\epsilon_4\epsilon_4)R(\Omega_+)^2-t_4t_4H^2R(\Omega_+)-\ft1{12}\epsilon_5\epsilon_5H^2R(\Omega_+)+\cdots\Bigr).
\label{eq:d6tree}
\end{equation}
Here $\alpha$ is inversely proportional to the volume of K3, and the $t_4$ tensor is taken to be $t_{4\,\mu_1\nu_1\mu_2\nu_2}=\eta_{\nu_2\mu_1}\eta_{\nu_1\mu_2}$ up to antisymmetrization in index pairs.  Our aim is to show that $\mathcal L_{\mathrm{tree}}^{\partial^4}$ vanishes, at least at the order of the cubic interactions that can be probed by the reduction of the quintic ten-dimensional action.  Actually, it is sufficient to demonstrate that it vanishes up to on-shell field redefinitions.  Thus we allow the use of the six-dimensional equations of motion
\begin{equation}
\nabla^\mu H_{\mu\alpha\beta}=2H_{\mu\alpha\beta}\partial^\mu\phi,\qquad
R_{\mu\nu}-\ft14H^2_{\mu\nu}=-2\nabla_\mu\nabla_\nu\phi,\qquad R=-4\partial\phi^2+\ft5{12}H^2,
\label{eq:eom0}
\end{equation}
as well as integration by parts in the action.

We begin with the quartic term
\begin{equation}
\mathcal L_{R^4}=\sqrt{-g}e^{-2\phi}\alpha(t_4t_4-\ft14\epsilon_4\epsilon_4)R(\Omega_+)^2,
\label{eq:Ltre}
\end{equation}
where
\begin{align}
t_4t_4R(\Omega_+)^2&= R_{\mu\nu}{}^{\alpha\beta}(\Omega_+)R^{\mu\nu}{}_{\alpha\beta}(\Omega_+),\nn\\
\ft14\epsilon_4\epsilon_4R(\Omega_+)^2&=R_{\mu\nu}{}^{\alpha\beta}(\Omega_+)R_{\alpha\beta}{}^{\mu\nu}(\Omega_+)-4R_\mu{}^\alpha(\Omega_+)R_\alpha{}^\mu(\Omega_+)+R(\Omega_+)^2.
\end{align}
The curvature with torsion is given by
\begin{equation}
R_{\mu\nu}{}^{\alpha\beta}(\Omega_+)=R_{\mu\nu}{}^{\alpha\beta}+\ft12(\nabla_\mu H_\nu{}^{\alpha\beta}-\nabla_\nu H_\mu{}^{\alpha\beta})+\ft14(H_\mu{}^{\alpha\rho}H_{\nu\rho}{}^\beta-H_\nu{}^{\alpha\rho}H_{\mu\rho}{}^\beta),
\end{equation}
and the Ricci contractions are
\begin{align}
R_\mu{}^\alpha(\Omega_+)&\equiv R_{\mu\rho}{}^{\alpha\rho}(\Omega_+)=R_\mu{}^\alpha-\ft12\nabla^\rho H_{\rho\mu}{}^\alpha-\ft14H_{\mu\rho\sigma}H^{\alpha\rho\sigma},\nn\\
R(\Omega_+)&\equiv R_\mu{}^\mu(\Omega_+)=R-\ft14H_{\mu\nu\rho}H^{\mu\nu\rho}.
\label{eq:RicO+}
\end{align}

As expected, the Riemann-squared terms cancel when the $t_4t_4$ and $\epsilon_4\epsilon_4$ terms are combined.  However, there is a subtlety when considering the torsionful case, as the order of the index contractions is different between the terms.  In particular, we use the identity
\begin{equation}
R_{\mu\nu}{}^{\alpha\beta}(\Omega_+)=R^{\alpha\beta}{}_{\mu\nu}(\Omega_-),
\end{equation}
which is a consequence of Bianchi, $dH=0$, to rewrite the Euler combination as
\begin{equation}
\ft14\epsilon_4\epsilon_4R(\Omega_+)^2=R_{\mu\nu}{}^{\alpha\beta}(\Omega_+)R^{\mu\nu}{}_{\alpha\beta}(\Omega_-)-4R_\mu{}^\alpha(\Omega_+)R^\mu{}_\alpha(\Omega_-)+R(\Omega_+)^2.
\end{equation}
As a result, we find
\begin{equation}
(t_4t_4-\ft14\epsilon_4\epsilon_4)R(\Omega_+)^2=2\nabla_\mu H_\nu{}^{\alpha\beta}\nabla^\nu H^\mu{}_{\alpha\beta}+4R_\mu{}^\alpha(\Omega_+)R^\mu{}_\alpha(\Omega_-)-R(\Omega_+)^2,
\label{eq:tteesum}
\end{equation}
where we have made use of $dH=0$ and $R_{\mu[\nu\alpha\beta]}=0$.  As we see, the Riemann-squared term cancels, but we are left with a non-trivial $(\nabla H)^2$ term as well as Ricci-like terms.

To proceed, we remove the derivatives acting on $H$ by a combination of integration by parts and the $H$ equation of motion.  We furthermore use the Einstein and $H$ equations of motion to rewrite the Ricci terms (\ref{eq:RicO+}) as
\begin{equation}
R_{\mu\nu}(\Omega_+)=-2\nabla_\mu\nabla_\nu\phi-H_{\mu\nu\lambda}\partial^\lambda\phi,\qquad
R(\Omega_+)=-2\square\phi.
\end{equation}
Inserting the above expressions into (\ref{eq:Ltre}) then gives
\begin{equation}
\mathcal L_{R^4}=\sqrt{-g}e^{-2\phi}\alpha[4R^{\mu\nu\rho\sigma}H_{\mu\rho\alpha}H_{\nu\sigma}{}^\alpha-2R^{\mu\nu}H^2_{\mu\nu}-4H^2_{\mu\nu}\nabla^\mu\nabla^\nu\phi
+16(\nabla_\mu\nabla_\nu\phi)^2-4H^2_{\mu\nu}\partial^\mu\phi\partial^\nu\phi-4(\Box\phi)^2].
\label{eq:LR4part}
\end{equation}
The first term is an irreducible three-point contact term which should not be present in the six-dimensional tree-level effective action.  Therefore it must be canceled by the reduction of the $H^2R^3$ terms in (\ref{eq:d6tree}) that we have yet to consider.

The additional $H^2R^3$ terms can be written explicitly as
\begin{equation}
t_4t_4H^2R(\Omega_+)=H^{\mu\nu\alpha}H^{\rho\sigma}{}_\alpha R_{\mu\nu\rho\sigma}(\Omega_+)=2R^{\mu\nu\rho\sigma}H_{\mu\rho\alpha}H_{\nu\sigma}{}^\alpha-\ft14H^4,
\end{equation}
and
\begin{align}
\epsilon_4\epsilon_4H^2R(\Omega_+)&=-\epsilon_{\alpha\mu_0\mu_1\cdots\mu_4}\epsilon^{\alpha\nu_0\nu_1\cdots\nu_4}H^{\mu_1\mu_2}{}_{\nu_0}H_{\nu_1\nu_2}{}^{\mu_0}R^{\mu_3\mu_4}{}_{\nu_3\nu_4}(\Omega_+)\nn\\
&=4\left[6R^{\mu\nu\rho\sigma}H_{\mu\rho\alpha}H_{\nu\sigma}{}^\alpha-6R^{\mu\nu}H^2_{\mu\nu}+RH^2-\ft12H^4-\ft12(H^2_{\mu\nu})^2-\ft14(H^2)^2\right],
\end{align}
where we have defined
\begin{equation}
H^4\equiv H_{\mu\nu\rho}H^{\mu ab}H^{\nu bc}H^{\rho ca}.
\end{equation}
As a result, the first term in (\ref{eq:LR4part}) is indeed canceled, and we end up with
\begin{align}
\mathcal L_{\mathrm{tree}}^{\partial^4}&=\sqrt{-g}e^{-2\phi}\alpha\bigl[-\ft13RH^2-4H^2_{\mu\nu}\nabla^\mu\nabla^\nu\phi+\ft16(H^2_{\mu\nu})^2+\ft5{12}H^4+\ft1{12}(H^2)^2\nn\\
&\kern6em+16(\nabla_\mu\nabla_\nu\phi)^2-4H^2_{\mu\nu}\partial^\mu\phi\partial^\nu\phi-4(\Box\phi)^2\bigr].
\end{align}
This effective four-derivative action still appears to have three-point interactions.  However, these can be pushed to higher order using a combination of equations of motion and integration by parts.  The final result is
\begin{equation}
\mathcal L_{\mathrm{tree}}^{\partial^4}=\sqrt{-g}e^{-2\phi}\alpha[\ft5{12}H^4+\ft16(H^2_{\mu\nu})^2+\ft{19}{36}(H^2)^2-8H^2_{\mu\nu}\partial^\mu\phi\partial^\nu\phi+\ft83H^2\partial\phi^2-16(\partial\phi^2)^2].
\label{eq:tree4}
\end{equation}
This is now written purely in terms of four-point contact interactions in six dimensions.  Of course, the entire set of tree-level four-derivative couplings in the gravity sector ought to vanish.  However, as these lift to six-point terms in ten dimensions, they remain unconstrained at the level of the five-point function that we have computed in (\ref{eq:treeea}).  Nevertheless, the presence of these terms directly indicates that the quintic effective action is incomplete and additional kinematical structures must necessarily be present that can only be probed at the level of the six-point function and beyond.

Until now, we have evaded the distinction between IIA and IIB compactifications on K3 as the NSNS fields are identical.  In ten dimensions, the NSNS fields can be viewed as the bosonic sector of a $\mathcal N=1$ theory that is extended to either IIA or IIB by the RR fields.  The compactification of the NSNS sector on K3 then results in a $(1,0)$ theory in six dimensions comprising a supergravity multiplet with bosonic fields $(g_{\mu\nu},b_{\mu\nu}^+)$, a $(1,0)$ tensor multiplet with bosonic fields $(b_{\mu\nu}^-,\phi)$ and 20 $(1,0)$ hypermultiplets with four scalars in each.  The hypermatter originates from the K3 moduli along with the ten-dimensional $B$-field compactified on two-cycles.  Note, however, that we only focus on the reduction to six-dimensional $(g_{\mu\nu},b_{\mu\nu},\phi)$, and hence will ignore the hypermatter couplings.

As noted above, the tree-level NSNS couplings of type II theory on K3 vanish at the four-derivative level.  (To see this from the reduction of the effective action requires knowledge of the six-point function in ten dimensions, which we have not computed.)  This indicates that neither the $(1,0)$ gravity multiplet nor the $(1,0)$ tensor multiplet receives any tree level four-derivative couplings in six dimensions.  However, the situation for one-loop couplings is rather different, and to discuss that we have to consider IIA and IIB reductions separately.

\subsection{Type IIA on K3}
\label{IIA/K3}

The compactification of Type IIA theory on K3 gives rise to $(1,1)$ supergravity coupled to 20 $(1,1)$ vector multiplets.  In particular, the RR fields can be thought of as adding a $(1,0)$ gravitino multiplet of opposite chirality as well as 20 $(1,0)$ vectors that combine with the $(1,0)$ hypermultiplets to fill out the $(1,1)$ vector multiplets.  What this indicates is that the six-dimensional NSNS fields $(g_{\mu\nu},b_{\mu\nu},\phi)$ reside in a single $(1,1)$ graviton multiplet, so whatever we obtain in the compactification of the IIA NSNS fields will provide information on the four-derivative couplings of the graviton multiplet.

At tree level, a direct reduction of the quintic action gives the six-dimensional couplings in (\ref{eq:tree4}).  However, as discussed above, we expect these couplings to be exactly cancelled by the addition of six-point terms in the ten-dimensional action.  Turning now to the one-loop level, the starting point is the one-loop effective Lagrangian, (\ref{eq:loopea}), with the top sign chosen for IIA theory, which can be reduced on K3 to give
\begin{equation}
\mathcal L_{\mathrm{loop}}^{\partial^4}=\sqrt{-g}\beta\Bigl((t_4t_4+\ft14\epsilon_4\epsilon_4)R(\Omega_+)^2+\ft16\epsilon_5\epsilon_5H^2R(\Omega_+)+\cdots+4 B_2\wedge\left[\tr R(\Omega_+)^2\right]_{\mbox{even in }B_2}\Bigr),
\end{equation}
where $\beta$ is a loop constant inversely related to the volume of K3.  Of course, we actually know more than this, as the one-loop effective action was previously obtained by dualizing the Heterotic tree-level four-derivative terms and by direct computation of a string four-point amplitude on the K3 orbifold \cite{Liu:2013dna}.  With a suitable normalization, the full NSNS result is given by
\begin{equation}
\mathcal L_{\mathrm{loop}}^{\partial^4}=\sqrt{-g}\fft{\alpha'}{16}\Bigl((t_4t_4+\ft14\epsilon_4\epsilon_4)R(\Omega_+)^2+\ft16\epsilon_5\epsilon_5H^2R(\Omega_+)+\ft1{36}\epsilon_4\epsilon_4H^4+4 B_2\wedge\left[\tr R(\Omega_+)^2\right]_{\mbox{even in }B_2}\Bigr).
\label{eq:iiaea}
\end{equation}
The additional $\epsilon_4\epsilon_4H^4$ term (with precise definition given in \cite{Liu:2013dna} after mapping $\epsilon_4\epsilon_4H^4\to-\fft12\epsilon_6\epsilon_6H^4$) can originate naturally from a $\epsilon_{8}\epsilon_{8}H^4R(\Omega_+)^2$ coupling in ten dimensions.  The addition of the RR fields will extend this effective Lagrangian to a complete set of four-derivative self couplings of the $(1,1)$ supergravity multiplet at one-loop level.  The full set of couplings can also be obtained by dualizing the heterotic theory reduced on $T^4$ while keeping the six-dimensional graviphotons that were not considered in \cite{Liu:2013dna}.

\subsection{Type IIB on K3}

We now turn to Type IIB theory on K3, where the RR sector extends the $(1,0)$ theory into $(2,0)$ theory in six dimensions.  Here, the $(1,0)$ supergravity and tensor multiplet from the NSNS sector complete into a $(2,0)$ supergravity and tensor multiplet.  In addition, the 20 $(1,0)$ hypermultiplets are completed into $(2,0)$ tensor multiplets, thus yielding $(2,0)$ supergravity coupled to 21 tensor multiplets in all.  Since we have only considered the NSNS fields, we restrict our focus to $(g_{\mu\nu},b_{\mu\nu},\phi)$ in six dimensions, corresponding to the supergravity multiplet and the tensor multiplet that is singled out by having $b_{\mu\nu}^-$ originating from the NSNS sector.  Nevertheless, in contrast to the IIA case, this is sufficient to provide information on the coupling of gravity and tensor multiplets, and not just on the gravity multiplet alone.

We start by noting the one-loop couplings for IIB theory on K3 can be obtained from (\ref{eq:iiaea}) by flipping the sign of the $\epsilon_4\epsilon_4$ and  $\epsilon_5\epsilon_5$ terms and by suitable modification in the CP-odd sector \cite{Liu:2013dna}
\begin{equation}
\mathcal L_{\mathrm{loop}}^{\partial^4}=\sqrt{-g}\beta\Bigl((t_4t_4-\ft14\epsilon_4\epsilon_4)R(\Omega_+)^2-\ft16\epsilon_5\epsilon_5H^2R(\Omega_+)-\ft1{36}\epsilon_4\epsilon_4H^4+4 B_2\wedge\left[\tr R(\Omega_+)^2\right]_{\mbox{odd in }B_2}\Bigr).
\end{equation}
Before proceeding, note that the purely gravitational terms take the form
\begin{equation}
(t_4t_4-\ft14\epsilon_4\epsilon_4)R^2=R_{\mu\nu\rho\sigma}^2-(R_{\mu\nu\rho\sigma}^2-4R_{\mu\nu}^2+R^2)=4R_{\mu\nu}^2-R^2,
\end{equation}
which vanishes on a Ricci-flat background that is consistent with setting $B_2=0$ and a constant dilaton.  This suggests that there is significant cancellation in the gravity sector of IIB theory reduced on K3, and we will see that this is in fact the case.


For the full set of NSNS fields, we start with the CP-even sector and note that the combination $(t_4t_4-\ft14\epsilon_4\epsilon_4)R(\Omega_+)^2$ was already worked out at tree level in (\ref{eq:tteesum}).  Again, we can simplify this using on-shell integration by parts.  However, at one-loop this proceeds without the $e^{-2\phi}$ tree-level factor.  The result is
\begin{align}
(t_4t_4-\ft14\epsilon_4\epsilon_4)R(\Omega_+)^2&=2\nabla_\mu H_\nu{}^{\alpha\beta}\nabla^\nu H^\mu{}_{\alpha\beta}+4R_\mu{}^\alpha(\Omega_+)R^\mu{}_\alpha(\Omega_-)-R(\Omega_+)^2,\nn\\
&=4R^{\mu\nu\rho\sigma}H_{\mu\rho\alpha}H_{\nu\sigma}{}^\alpha-2R^{\mu\nu}H^2_{\mu\nu}+4R_\mu{}^\alpha(\Omega_+)R^\mu{}_\alpha(\Omega_-)-R(\Omega_+)^2+2(\nabla^\mu H_{\mu\alpha\beta})^2\nn\\
&=4R^{\mu\nu\rho\sigma}H_{\mu\rho\alpha}H_{\nu\sigma}{}^\alpha-2R^{\mu\nu}H^2_{\mu\nu}+4(R_{\mu\nu}-\ft14H^2_{\mu\nu})^2-(R-\ft14H^2)^2+(\nabla^\mu H_{\mu\alpha\beta})^2.
\label{eq:tteeR}
\end{align}
This may be combined with
\begin{equation}
-\ft16\epsilon_5\epsilon_5H^2R(\Omega_+)-\ft1{36}\epsilon_4\epsilon_4H^4=-4R^{\mu\nu\rho\sigma}H_{\mu\rho\alpha}H_{\nu\sigma}{}^\alpha+4R^{\mu\nu}H^2_{\mu\nu}-\ft23RH^2+\ft1{18}(H^2)^2-\ft13H^4.
\end{equation}
to give the effective Lagrangian in the CP-even sector
\begin{align}
\mathcal L_{\textrm{CP-even}}&=\sqrt{-g}\beta\Bigl(2R^{\mu\nu}H_{\mu\nu}^2+4(R_{\mu\nu}-\ft14H_{\mu\nu}^2)^2-(R+\ft1{12}H^2)^2-\ft13H^4+(\nabla^\mu H_{\mu\alpha\beta})^2\Bigr)\nn\\
&=\sqrt{-g}\beta\Bigl(2R^{\mu\nu}H^2_{\mu\nu}+16(\nabla_\mu\nabla_\nu\phi)^2-(4\partial\phi^2-\ft12H^2)^2-\ft13H^4+4H^2_{\mu\nu}\partial^\mu\phi\partial^\nu\phi\Bigr),
\label{eq:CPe}
\end{align}
where the second line is obtained by substituting in the lowest order equations of motion.


We now turn to the CP-odd contribution, which takes the form \cite{Liu:2013dna}
\begin{align}
\mathcal L_{\textrm{CP-odd}}&=4\beta B\wedge\tr[R(\Omega_+)\wedge R(\Omega_+)]_{\mathrm{odd\ in\ }B_2}\nn\\
&=4\beta B\wedge\tr[d\mathcal H\wedge(R+\ft14\mathcal H^2)]\nn\\
&=4\beta B\wedge d\tr[\mathcal H\wedge(R+\ft1{12}\mathcal H^2)]=4\beta H\wedge\tr[\mathcal H\wedge(R+\ft1{12}\mathcal H^2)].
\label{eq:CPodd}
\end{align}
Note that the one-form $\mathcal H^{ab}=H_\mu{}^{ab}dx^\mu$ is what shows up in the torsionful connection $\Omega_+=\Omega+\fft12\mathcal H$.  The $H\wedge\tr\mathcal H\wedge R$ term involves the Riemann tensor.  However, as demonstrated in \cite{Liu:2013dna}, it can be rewritten purely kinematically in terms of the Ricci tensor.  Then, in component notation, (\ref{eq:CPodd}) becomes
\begin{align}
\mathcal L_{\textrm{CP-odd}}&=\sqrt{-g}4\beta\fft1{3!}\epsilon^{\mu_1\mu_2\mu_3\mu_4\mu_5\mu_6}H_{\mu_1\mu_2\mu_3}[\ft1{2!}H_{\mu_4\mu_5}{}^a R_{\mu_6}^a-\ft1{12}H_{\mu_4}{}^{ab}H_{\mu_5}{}^{bc}H_{\mu_6}{}^{ca}]\nn\\
&=-\sqrt{-g}4\beta(*H)^{\mu\nu\rho}[\ft1{2!}H_{\mu\nu}{}^a R_{\rho}^a-\ft1{12}H_{\mu}{}^{ab}H_{\nu}{}^{bc}H_{\rho}{}^{ca}]\nn\\
&=-\sqrt{-g}\beta\Bigl(2R^{\mu\nu}(H^{(+)}_{\mu\alpha\beta}-H^{(-)}_{\mu\alpha\beta})H_\nu{}^{\alpha\beta}-\ft13(H^{(+)}-H^{(-)})H^3\Bigr),
\label{eq:CPo}
\end{align}
where the last line is obtained by rewriting $H$ in terms of its self dual and anti-self dual components $H=H^{(+)}+H^{(-)}$, so that $*H=H^{(+)}-H^{(-)}$.  Note that in the last term we are using a short-hand notation for the non-factorized $H^4$ combination.

We are now ready to combine the CP-even and CP-odd couplings, (\ref{eq:CPe}) and (\ref{eq:CPo}), respectively, with the result
\begin{equation}
\mathcal L_{\mathrm{loop}}^{\partial^4}=\sqrt{-g}\beta\Bigl(4R^{\mu\nu}H_{\mu\alpha\beta}^{(-)}H_\nu{}^{\alpha\beta}+16(\nabla_\mu\nabla_\nu\phi)^2-(4\partial\phi^2-\ft12H^2)^2-\ft23H^{(-)}H^3+4H^2_{\mu\nu}\partial^\mu\phi\partial^\nu\phi\Bigr).
\end{equation}
This expression can be simplified through a combination of integration by parts and application of the on-shell equations of motion, with the result
\begin{equation}
\mathcal L_{\mathrm{loop}}^{\partial^4}=\sqrt{-g}\beta\Bigl(-\ft23H^{(-)}H^3+H^{(-)}_{\mu\alpha\beta}H_\nu{}^{\alpha\beta}H^{2\,\mu\nu}-\ft5{36}(H^2)^2-8H^{(-)}_{\mu\alpha\beta}H_\nu{}^{\alpha\beta}\nabla^\mu\nabla^\nu\phi+16(\partial\phi^2)^2\Bigr).
\end{equation}
The $H^{(-)}H\nabla\nabla\phi$ term can be simplified by decomposing $H^{(-)}_{\mu\alpha\beta}H_\nu{}^{\alpha\beta}=H^{(-)\,2}_{\mu\nu}+H^{(-)}_{\mu\alpha\beta}H_\nu{}^{(+)\,\alpha\beta}$ and using the identity
\begin{equation}
H^{(-)\,\alpha\beta}_{(\mu}H_{\nu)\alpha\beta}^{(+)}=\ft16g_{\mu\nu}H^{(-)}_{\rho\alpha\beta}H^{(+)\,\rho\alpha\beta}=\ft1{12}g_{\mu\nu}H^2.
\label{eq:Hpmmn}
\end{equation}
This results in an $H^2\Box\phi$ term which is then simplified using the equations of motion.  The result is
\begin{align}
\mathcal L_{\mathrm{loop}}^{\partial^4}&=\sqrt{-g}\beta\Bigl(-\ft23H^{(-)}H^3+H^{(-)}_{\mu\alpha\beta}H_\nu{}^{\alpha\beta}H^{2\,\mu\nu}-\ft1{12}(H^2)^2
-8H^{(-)\,2}_{\mu\nu}\nabla^\mu\nabla^\nu\phi-\ft43H^2\partial\phi^2+16(\partial\phi^2)^2\Bigr)\nn\\
&=\sqrt{-g}\beta\Bigl(\ft43H^{(-)\,4}-8H^{(-)\,2}_{\mu\nu}\nabla^\mu\nabla^\nu\phi-\ft43H^2\partial\phi^2+16(\partial\phi^2)^2\Bigr),
\label{eq:tensorfin}
\end{align}
where the second line is obtained by fully decomposing the $H^4$ terms into self dual and anti-self dual components.

As written, the $H^{(-)\,2}\nabla\nabla\phi$ term in (\ref{eq:tensorfin}) appears to represent a contact three-point interaction.  However, as demonstrated in Appendix~\ref{sec:Hsimp}, it can be on-shell manipulated to give
\begin{equation}
H^{(-)\,2}_{\mu\nu}\nabla^\mu\nabla^\nu\phi=-2H^{(-)\,2}_{\mu\nu}\partial^\mu\phi\partial^\nu\phi-\ft16H^2\partial\phi^2.
\end{equation}
As a result, we find the $(2,0)$ loop couplings
\begin{equation}
\mathcal L_{\mathrm{loop}}^{\partial^4}=\sqrt{-g}\beta\Bigl(\ft43H^{(-)\,4}+16H^{(-)\,2}_{\mu\nu}\partial^\mu\phi\partial^\nu\phi+16(\partial\phi^2)^2\Bigr).
\label{eq:final}
\end{equation}
Note that this only involves $(b_{\mu\nu},\phi)$ which are the bosonic components of the $(1,0)$ tensor multiplet that is extended to a $(2,0)$ tensor with the inclusion of RR fields that we have ignored.  Moreover, it is a quartic four-derivative tensor self-coupling that obeys the requirements of $(2,0)$ supersymmetry as outlined in \cite{Lin:2015dsa}.  In fact, the only quartic four-derivative couplings that are allowed are between $(2,0)$ tensors, and this is consistent with the absence of any gravity multiplet fields in (\ref{eq:final})%
\footnote{Note that tree-level quartic couplings among $(2,0)$ tensors are also expected to be present.  In fact, the distinction between tree and loop is somewhat lost when considering the $O(5,21)$ structure of the $(2,0)$ theory, and generically all 21 tensor multiplets will have moduli-dependent quartic couplings.  However, the particular tensor multiplet obtained from the NSNS reduction has vanishing tree-level self-couplings, as we have seen above.}.
The explicit supersymmetrisation of various one-loop $R^2$ terms in a general $(1,0)$ setup was carried out in \cite{Novak:2017wqc, Butter:2018wss}. Similar calculations for the quartic interactions of $(1,0)$ and $(2,0)$ tensor multiplets should be of considerable interest.

The gravity multiplet fields that are accessible from the NSNS sector are $(g_{\mu\nu},b_{\mu\nu}^{(+)})$.  The vanishing of $R_{\mu\nu\rho\sigma}^2$ and $R_{\mu\nu\rho\sigma}H^{\mu\nu}{}_\alpha H^{\rho\sigma\alpha}$ terms suggest that the $(\mbox{graviton})^4$ and mixed $(\mbox{graviton})^2(\mbox{tensor})^2$ couplings are not present.  However, the full absence of such couplings is only verified after a somewhat intricate splitting of $H$ into its self dual and anti-self dual components.  In particular, a non-trivial IIB CP-odd term, (\ref{eq:CPodd}), is required to obtain the proper decoupling of the $(2,0)$ gravity multiplet at the four-derivative level.  This provides direct evidence that ten-dimensional IIB theory has a one-loop eight derivative CP-odd term, even though it is completely absent at tree level.

\subsection{Reduction on CY threefolds}

In Calabi-Yau reductions the order of derivatives appearing in the lower dimensional action is controlled by the power of the CY Riemann tensors appearing in the internal integrals. Hence for the threefold reduction to four dimensions the four-derivative couplings discussed in the previous subsections are not the lowest order contributions. These lowest order two-derivative corrections manifest themselves in the corrections to the moduli space metrics proportional to the CY Euler number. The correction to the vector multiplet moduli space  $G_{vv}$ is at tree-level and to the hypemultiplet moduli space metric  $G_{hh}$ at one loop \cite{Antoniadis:1997eg, Antoniadis:2003sw}. These calculations are done directly in the CY background and did not rely on a reduction of ten-dimensional $R^4$ terms. However they predict the result of such a reduction, and the corrected four-dimensional effective action in string frame should be given by 
\begin{align}
\label{eq:ef4d}
S&=\fft1{2\kappa_4^2}\int d^4x\sqrt{-g}
\Bigl[ \left( (1 + {\chi_T} ) e^{-2\phi_4} - \chi_1\right) R_{(4)} \nonumber\\
&\quad + \left( (1 - {\chi_T}) e^{-2\phi_4} - \chi_1\right) G_{vv} (\partial v)^2 +  \left((1 + {\chi_T}) e^{-2\phi_4} + \chi_1\right) G_{hh} (\partial h)^2 \Bigr],
\end{align}
where $\chi_T$ and $\chi_1$ are defined in terms of the CY Euler number $\chi$ and volume $V_6$ as:
\begin{align}
\chi_T &= \frac{2 \zeta(3) (2 \pi \alpha')^3 }{V_6} \, \chi, \nn \\
\chi_1 &= \frac{4 \zeta(2) }{(2\pi)^3} \, \chi.
\end{align}

The first step in obtaining \eqref{eq:ef4d} via reduction is  working out the $\sim (\alpha')^3$ corrections to the CY background. The well-known result for the dilaton and Riemann tensor is
\begin{align}
\label{eq:cora3}
\phi &=\phi_0+\fft{\alpha'^3}{3\cdot2^{8}}\left(\zeta(3)+\fft{(2\pi)^2}{12}e^{2\phi_0}\right)E_6, \nn \\
R_{i\bar j} &= -\fft{\alpha'^3}{3\cdot2^{6}}\left(\zeta(3)+\fft{(2\pi)^2}{12}e^{2\phi_0}\right)
\partial_i\partial_{\bar j}E_6,
\end{align}
with the expansion around the constant dilaton $\phi_0$ and internal Ricci-flat metric. Here $E_6$ is the Euler density of the CY manifold. The action \eqref{eq:ef4d} should be a result of combining the  reduction of ten-dimensional classical action using \eqref{eq:cora3} and the $R^4$ terms evaluated on a classical CY metric. This is rather involved and requires a number of non-trivial cancellations. This computation was carried out in \cite{Grimm:2017okk}, where it was pointed out that a tree-level $-2e^{-2\phi}t_8t_8H^2R^3$ coupling would be necessary for these cancellations to happen.  The tree-level result, (\ref{eq:treeea}), indeed confirms the presence of such a term.  The other terms in (\ref{eq:treeea}), namely $-\frac16 e^{-2\phi}\epsilon_{9}\epsilon_{9}H^2R^3$ and $e^{-2\phi} \sum d_i H^2 \cdot {\tilde Q}^i$, do not contribute to the reduction.  Similarly, there is no contribution from the one-loop $-\frac13\epsilon_{9}\epsilon_{9}H^2R^3$ term in (\ref{eq:loopea}).

The $R^2$ terms in reductions on a Calabi-Yau manifold $X$ at linearised level were discussed in \cite{Antoniadis:1997eg} and work out in a very similar fashion to $K3$ reductions with the following replacement:
\bea
\frac13 \int_{K3} p_1 =  -16 \,\, &\mapsto& \frac13 \int_X \omega_I \wedge p_1(TX) = \alpha_ I , \nn \\
\frac{1}{32\pi^2} \int_{K3} (Riem)^2 = 24 \,\, &\mapsto& \,\,  \frac{1}{32\pi^2} \int_X (Riem)^2 = \alpha_I t^I  ,
\eea
where $\omega^I \in H^{1,1}(X, \mathbb{Z})$ with $I = 1, \ldots, h^{1,1}(X)$, $\alpha^I$ is a set of topological numbers, and $t^I$ are the K\"ahler moduli. In full agreement with the $\cN = 2$ special geometry, the only non-vanishing $R^2$ couplings appear  at one loop in the type IIA reductions and are of the form
\beq
\alpha^I \Bigl[ u^I \tr R\wedge R + t^I R_{\m\n\r\l} R^{\m \n\r \l} \Bigr],
\eeq
where $u^I$ are the $h^{1,1}(X)$ moduli coming from the $B$-field. From the point of view of supersymmetry, this coupling does not need the $B$-field that is no longer in the gravity multiplet. However four-derivative couplings are expected in both vector and hyper multiplets. Moreover, starting from four-derivatives there are mixed couplings involving vectors and hypers, and starting from six derivatives, mixed couplings also involving the gravity and matter multiplets. The stringy origin of these couplings has been studied only for the reduction of type IIA one-loop couplings \cite{Katmadas:2013mma} since these are the only ones that lift to M-theory. Their better understanding on the type IIB side together with the full $SO(5)$ completions of six-dimensional $(2,0)$ theories should be useful in finding a completion of the higher-derivative couplings in string theory with RR fields included.

\section{Symmetries and dualities}
\label{sec:susy}

At the linear level, the structure of the $R^4$ couplings is very similar at the tree and loop level, and this has led to important observations on supersymmetry as well as $SL(2,\mathbb Z)$ duality of the IIB string.  However, as we have seen above, the non-linear kinematics is surprisingly different at tree-level and one-loop.  Thus both supersymmetry as well as $SL(2,\mathbb Z)$ invariance of the IIB couplings need to be revisited in light of this difference.

\subsection{Supersymmetry}

Based on the linearised computations, and ignoring the dilaton (which is unimportant in the quartic effective action), the tree-level and one-loop $R^4$ contributions are often grouped into two $\mathcal{N}=1$ superinvariants:
\begin{align}
\label{superinv}
J_0(\Omega) &= \left(t_8 t_8 - \ft14 \epsilon_{8} \epsilon_{8}\right)R^4  , \nn \\
J_1(\Omega) &= t_8 t_8 R^4  -  \ft14 \epsilon_{10} t_8 B R^4.
\end{align}
The IIA and IIB tree-level couplings are given by $J_0$, while at one-loop IIB is once more given by $J_0$, but IIA is given by $2J_1 - J_0$. Note that the IIA action has a CP-odd term $\ft14 \epsilon_{10} t_8 B R^4$ at one-loop, while IIB does not.

The disparity between the tree-level and one-loop couplings at the non-linear level, as well as the appearance of a CP-odd coupling even in $B$-fields in the IIB action, makes it impossible to describe the full action by only two $R^4$ superinvariants.  At the non-linear level, we expect there are two independent one-loop invariants which may be taken to be
\begin{align}
\label{superinv1}
\JJ_t(\Omega_+) &= t_8 t_8 R^4 (\Omega_+)  -  \ft14 t_8 \epsilon_{10} B R^4 (\Omega_+) , \nn \\
\JJ_{\epsilon} (\Omega_+) &=   \ft14 \epsilon_8 \epsilon_8 R^4 (\Omega_+) -  \ft14  \epsilon_{10} t_8 B R^4 (\Omega_+)  + \ft 13 \epsilon_{9} \epsilon_{9} H^2 R^3 (\Omega_+) - \ft{4}{9}  \epsilon_{9}\epsilon_{9}H^2 (\nabla H)^2 R + \cdots.
\end{align}
These are expected to be ${\mathcal N}=(1,0)$ invariants built from the NSNS fields, and supersymmetry is essentially acting by swapping $t_8\leftrightarrow\epsilon_{10}$ only on one side of the string worldsheet.  Construction of ${\mathcal N}=2$ invariants then corresponds to the addition of an $\mathcal N=(1,0)$ chiral gravitino or anti-chiral gravitino multiplet depending on the completion to IIA or IIB.  In any case, the IIA/IIB one-loop terms are then given by $\JJ_{\mathrm{IIA/IIB}} = \JJ_t \pm \JJ_{\epsilon}$. Note that the parity of the RR-fields and the relative sign between $\JJ_t$ and $\JJ_{\epsilon}$ (GSO projection) are correlated. So at the level of ${\mathcal N}=2$ we should again recover a single  one-loop invariant. Obviously ${\mathcal N} = (1,1) \,\, \mbox{and} \,\, (2,0)$ invariants are different from each other. At the four-point level, combinations of $\JJ_t$ and $\JJ_{\epsilon}$ reduce to the standard invariants \eqref{superinv}.

The tree-level action \eqref{eq:treeea} derived above is supposed to be independently supersymmetric and hence defines a third $\mathcal N=(1,0)$ supersymmetry invariant:
\begin{align}
\JJ_0(\Omega_+) &= e^{-2\phi}
\Bigl(t_8t_8R(\Omega_+)^4-\ft14\epsilon_{8}\epsilon_{8} R(\Omega_+)^4-2t_8t_8H^2R(\Omega_+)^3-\ft16\epsilon_{9}\epsilon_{9}H^2R(\Omega_+)^3\nn\\
&\kern4em+8\cdot4!\sum_id_iH^{\mu\nu\lambda}H^{\rho\sigma\zeta}\tilde Q^i_{\mu\nu\lambda\rho\sigma\zeta}+\cdots\Bigr).
\end{align}
This invariant can be promoted to either ${\mathcal N} = (1,1) \,\, \mbox{or} \,\, (2,0)$. At the non-linear level, the $e^{-2\phi}$ dilaton factor cannot be ignored and makes this invariant distinct from any of the one-loop invariants.  As an $\mathcal N=(1,0)$ invariant, we expect this to also appear in the tree-level heterotic action%
\footnote{Additional couplings to gauge fields in $\mathcal N=(1,0)$ vector multiplets may be required as well.}.
It would be interesting to see if this is the case.

\subsection{\texorpdfstring{$SL(2,\mathbb Z)$}{SL(2,Z)} invariance of \texorpdfstring{$R^4$}{R4} couplings}
\label{SL(2,Z)}

At the two-derivative level, the classical IIB action is invariant under $SL(2,\mathbb R)$ transformations, while the full theory is expected to be invariant under $SL(2,\mathbb Z)$ duality transformations.  Given the presence of additional $H^2R^3$  couplings with different kinematical structures at both tree and one-loop level, we now address the question of how they can be reconciled with $SL(2, \bZ)$ duality.

Since $SL(2, \bZ)$ invariance is more naturally investigated in the Einstein frame, we transform from the string frame, which was used above, to the Einstein frame by taking $g_{\mu\nu}^{(\mathrm{string})}=e^{\phi/2}g_{\mu\nu}^{(\mathrm{Einstein})}$.  Collecting the tree-level contribution from (\ref{eq:treeea}) and the one-loop contribution from (\ref{eq:loopea}), the perturbative part of the type IIB eight-derivative action up to $H^2R^3$ terms is given in the Einstein frame as
\begin{align}
\mathcal L_{\mathrm{IIB}}^{\partial^8} &= \alpha'^3 \sqrt{-g} \biggl[ \Bigl( e^{-\frac32 \phi}\fft{\zeta(3)}{3\cdot2^{11}} +  e^{\frac12 \phi} \fft{\pi^2}{9\cdot2^{11}} \Bigr)\left(t_8t_8R({\hat \Omega}_+)^4-\ft14\epsilon_{8}\epsilon_{8} R({\hat \Omega}_+)^4  \right)\nn \\
&\kern5em+e^{- \frac32\phi}\fft{\zeta(3)}{3\cdot2^{11}}e^{-\phi}\Bigl(-\ft16\epsilon_{9}\epsilon_{9}H^2R({\hat \Omega}_+)^3 -2t_8t_8H^2R({\hat \Omega}_+)^3+8\cdot4!\sum_id_iH^{\mu\nu\lambda}H^{\rho\sigma\zeta}\tilde Q^i_{\mu\nu\lambda\rho\sigma\zeta}+\cdots\Bigr) \nn\\
&\kern5em+ e^{\frac12 \phi} \fft{\pi^2}{9\cdot2^{11}}e^{-\phi}\left(-\ft13\epsilon_{9}\epsilon_{9}H^2R({\hat \Omega}_+)^3+\cdots\right) \biggr]\nn \\
&\qquad-\fft{(2\pi)^2}{3\cdot2^6} \alpha'^3   B_2\wedge\Bigl[\tr R(\hat\Omega_+)^4-\ft14(\tr R(\hat\Omega_+)^2)^2\Bigr]_{\mbox{\scriptsize odd in $B_2$ }}.
\label{eq:IIB}
\end{align}
The first line collects the terms that are kinematically the same at tree-level and one loop. The second and third lines contain respectively tree-level and one-loop terms that are different. Finally the last line is the one-loop CP-odd contribution. Note that in the Einstein frame the torsionful connection picks up a dilaton contribution
\beq
(\O_+)_{\m}{}^{\a\b} \,\, \longrightarrow \,\, ({\hat \Omega}_+)_{\m}{}^{\a\b} = \Omega_{\m}{}^{\a\b}+\ft12e^{-\phi/2} H_\mu{}^{\alpha\beta} +\ft12 e_{\m}^{[\alpha} e^{\beta]\rho}\partial_\rho \phi,
\eeq
where all quantities on the right-hand side are in the Einstein frame.  Note that the ellipses in (\ref{eq:IIB}) include terms of the form $H^2(\nabla H)^2R$ as well as terms at higher order and possible additional dilaton terms.  The additional factor of $e^{-\phi}$ in the middle two lines is associated with $H^2$ in the Einstein frame.

Except for the metric and RR five-form, the IIB fields transform nontrivially under $SL(2,\mathbb Z)$.  Hence the various terms in (\ref{eq:IIB}) must arrange themselves into combinations that restore the invariance.  A full investigation of how this works requires knowledge of the RR sector.  However, even without a complete picture of the RR couplings, we can go a long way towards exploring the $SL(2,\mathbb Z)$ structure of the action by introducing the standard definitions
\bea
\tau = C_0 + i e^{-\phi} \, , \quad  P_m = \fr{i}{2 \tau_2 } \na_m \t \quad \mbox{and} \quad G_3 = \frac{1}{\sqrt{\tau_2}}(F_3 - \tau H_3),
\eea
where $C_n$ and $F_{n+1}$ are respectively  RR potentials and their field strengths.  Note that the complex fields $P$ and $G$ transform with charges $+2$ and $+1$, respectively, under the local $U(1)$ symmetry of IIB theory.  In order to obtain $SL(2,\mathbb Z)$ invariant couplings, we must then multiply the contact terms built out of $R$, $P$ and $G$ by  appropriate $SL(2,\mathbb Z)$ covariant functions of opposite charge.

In the pure gravity sector, the $(t_8t_8-\fft14\epsilon_8\epsilon_8)R^4$ combination in the first line of (\ref{eq:IIB}) is naturally made invariant by introducing the function $f_0(\tau,\bar\tau)$, which is the $SL(2,\bZ)$-invariant, non-holomorphic Eisenstein series of weight $3/2$ 
\beq
f_0 (\t, \bar \t) = \mathcal E_{3/2}(\tau,\bar\tau) =  \sum_{ (m,n)\neq (0,0)} \fr{\tau_2^{3/2}}{|m + n \t |^3}.
\eeq
At large values of $\t_2$, this has the expansion
\beq
\label{f0expansion}
f_0 (\t, \bar \t) = 2 \z (3)\, \t_2^{3/2} \;+\; \fr{2\pi^2}{3}\, \tau_2^{-1/2}\; +\; \cO(e^{-\tau_2}),
\eeq
which generates the tree-level, one-loop and non-perturbative $R^4$ terms in the effective action \cite{Green:1997tv}.

In order to account for couplings with $P$ and $G$ that carry $U(1)$ charge, we note that  $f_0(\tau,\bar\tau)$ is part of a family of functions $f_k (\t, \bar \t)$ given by
\beq
f_k (\t, \bar \t) = \sum_{ (m,n) \neq (0,0)} \fr{ \t_2^{\fr32}}{( m + n \t )^{\fr32 + k } ( m + n \bar \t )^{\fr32-k}},
\label{fdef}
\eeq
which transform under $SL(2,\mathbb Z)$ as
\begin{equation}
f_k\left(\fft{a\tau+b}{c\tau+d},\fft{a\bar\tau+b}{c\bar\tau+d}\right)=\left(\fft{c\tau+d}{c\bar\tau+d}\right)^kf_k(\tau,\bar\tau),
\end{equation}
and which carries $U(1)$ charge $-2k$.  Note that $f_k(\tau,\bar\tau)$ satisfy the relations
\beq
\bar f_k = f_{-k} \, , \quad 
 \left( k + 2i \t_2 \fr{\pa}{\pa \t} \right) f_k  = \left( k +  \fr32  \right) f_{k+1}  \quad \mbox{and} \quad 
 \left( k + 2i \t_2 \fr{\pa}{\pa \bar \t} \right) f_k = \left( k - \fr32 \right) f_{k-1},
\label{frelation}
\eeq
from which it is possible to demonstrate that
\bea
D_m f_k = - \left( \fr32 + k \right) f_{k+1} P_m -  \left( \fr32 - k \right) f_{k-1} \bar P_m.
\label{derivfrelation}
\eea
Here
\begin{equation}
D_m=\nabla_m-iq\fft{\partial_m\tau_1}{2\tau_2}
\end{equation}
is the $U(1)$ covariant derivative acting on an expression with charge $q$.  Note that the large $\tau_2$ expansion of (\ref{fdef}) takes the form
\begin{equation}
f_k (\t, \bar \t) = 2 \z (3)\, \t_2^{3/2} \;+\; \fr{2\pi^2}{3(1-4k^2)}\, \tau_2^{-1/2}\; +\; \cO(e^{-\tau_2}).
\label{eq:fkexpansion}
\end{equation}
In particular, the tree and loop coefficients are now related by a $k$-dependent factor.

It was suggested in \cite{Kehagias:1997cq} that the quartic effective action obtained from the first line of (\ref{eq:IIB}) can be written in the form
\begin{equation}
\mathcal L_{\mathrm{4pt}}=\sum_{k=0}^4f_k(\tau,\bar\tau)W_{2k}^{(4)}(R,\nabla P,\nabla\bar P,\nabla G,\nabla\bar G)+c.c.,
\end{equation}
where $W_{2k}^{(4)}(\ldots)$ represents combinations of total charge $+2k$ that is quartic in the fields.  Note, however, that the $U(1)$ charge assignment is ambiguous without knowledge of the RR couplings.  For example, $H_3$ in the NSNS sector can be extended to either $G_3$ or $\bar G_3$, carrying opposite charges.  This ambiguity can be resolved  following a complete computation of the quartic action including the RR fields \cite{Policastro:2006vt,Policastro:2008hg}, with the result valid at the level of the four-point couplings
\begin{align}
\label{eq:sl2-4pt}
\cL_{\mathrm{4pt}}&=\fft{\alpha'^3}{3\cdot2^{12}}\sqrt{-g}\Bigl[f_0 (\t, \bar \t) \Bigl(R^4 + 6 R^2 (4|\na P|^2 + |\na G|^2) + 24 |\na P|^2 |\na G|^2  + \hat{\cO}_1((|\na P|^2)^2) + \hat{\cO}_2((|\na G|^2)^2) \nn \\
&\kern11em+ 12\left(R\nabla P(\nabla{\bar G})^2 + R\nabla{\bar P}(\nabla G)^2\right)\Bigr)\Bigr].
\end{align}
The terms here are written schematically, and a contraction with $t_8t_8-\ft14\epsilon_{8}\epsilon_{8}$ is implied except for the $\hat{\mathcal O}_1$ and $\hat{\mathcal O}_2$ terms, which do not have this kinematical structure.  Note that the last term in \eqref{eq:sl2-4pt}, which can be expanded as
\begin{equation}
R\nabla P(\nabla{\bar G})^2 + R\nabla{\bar P}(\nabla G)^2=2R (\nabla\nabla \phi) [ H^2 - (F_3)^2] + 4 R \nabla F_1 H_3 F_3,
\label{eq:fixed}
\end{equation}
is different from the interaction $R(\nabla P+\nabla\bar P)|\nabla G|^2$ carrying $U(1)$ charge $\pm2$ that appears in \cite{Policastro:2008hg}.  As mentioned in the introduction, the combination (\ref{eq:fixed}) preserves the local $U(1)$ symmetry like all the other quartic terms, in agreement with the results of \cite{Boels:2012zr, Green:2019rhz}. We have verified explicitly from the two NSNS--two RR amplitude of \cite{Policastro:2006vt} that there is a relative minus sign between $R (\nabla\nabla \phi) H^2$ and  $R (\nabla\nabla \phi) (F_3)^2$, thus confirming the $(\nabla G)^2$ as opposed to the $|\nabla G|^2$ form of the expression in (\ref{eq:fixed}).  We also note that while all purely NSNS expressions in IIB theory are all even in powers of $H$, as required by world-sheet parity under which $B \rightarrow - B$, this does not have to be the case for the mixed interactions. In particular due to $F_{2n+1} \rightarrow (-1)^n F_{2n+1}$ under world-sheet parity, interactions of the form $R \nabla F_1 H_3 F_3$ are perfectly consistent.

As pointed out in \cite{Policastro:2008hg}, the passage from $\phi$ and $H$ to $\t$ and $G$ also brings in new kinematic mixing between the NS and RR sectors. These new contraction structures have the property
\beq
{\hat \cO}_i((\na X)^4) = (t_8t_8-\ft14\epsilon_{8}\epsilon_{8})(\na X)^4,
\eeq
for $X$ being either of $\na \phi$, $F_1$, $H_3$ or $F_3$, but
\beq
{\hat \cO}_i((\na X)^2 (\na Y)^2) \neq (t_8t_8-\ft14\epsilon_{8}\epsilon_{8})(\na X)^2 (\na Y)^2,
\eeq
for different $X$ and $Y$.

The $\hat{\cO}_1((|\na P|^2)^2)$ term has been understood in terms of the F-theory lift, and can be compactly written using the familiar $t_8t_8$ and $\epsilon_8\epsilon_8$ structures by formally extending the range of indices and using a special metric for the contractions. Guessing the structure of these kinds of terms starting from NS expressions is not possible, and the incorporation of the RR sector is needed in order to fully determine the higher-point kinematics. We are not going to pursue this here, but will simply discuss the general $SL(2, \bZ)$ invariance of the IIB action.

Of course, the first line of (\ref{eq:IIB}) goes well beyond the quartic level and includes terms up to the level of the eight-point function, such as some combination of $H$ or $\nabla\phi$ to the eighth power.  These quintic and higher terms carry $U(1)$ charges up to $\pm16$, and hence would need to be multiplied by $f_k(\tau,\bar\tau)$ up to $k=\pm8$ to create $SL(2,\mathbb Z)$ invariants.  Again, the form of such terms can only be pinned down with information from the RR sector.  However, we note from (\ref{eq:fkexpansion}), that the relative factor between the tree and loop contributions will differ from that in the first line of (\ref{eq:IIB}) for all terms with $k\ne0$.  In fact, consistency with the NSNS sector demands that all terms obtained from the first line of (\ref{eq:IIB}) with $k\ne0$ must vanish when restricted to the NSNS sector.  This actually provides a rather strong constraint on the form of the $U(1)$ charged terms, and hence also on the RR sector couplings.  For example, the quartic action (\ref{eq:sl2-4pt}) could in principle have had terms up to $k=\pm4$ \cite{Kehagias:1997cq}, yet all terms with $k\ne0$ are not present as they would be inconsistent with having identical tree and loop kinematics in the NSNS sector.

We now consider the terms of the form $H^2R^3$ in the middle two lines of (\ref{eq:IIB}).  At quintic order, these terms can be completed to either $|G|^2R^3$ or $(G^2+\bar G^2)R^3$.  As a result, the $SL(2,\mathbb Z)$ invariant combination necessarily takes the form
\begin{align}
a_0f_0(\tau,\bar\tau)|G^2|R^3+a_1\left(f_1(\tau,\bar\tau)G^2R^3+c.c.\right)~&\underset{\text{NSNS}}\longrightarrow~
\bigl(a_0f_0(\tau,\bar\tau)-a_1\left(f_1(\tau,\bar\tau)+f_{-1}(\tau,\bar\tau)\right)\bigr)e^{-\phi}H^2R^3\nn\\
&\longrightarrow~\left(2\zeta(3)e^{-3\phi/2}(a_0-2a_1)+\fft{2\pi^2}3e^{\phi/2}(a_0+\ft23a_1)\right)e^{-\phi}H^2R^3,
\label{eq:a0a1}
\end{align}
where we made use of (\ref{eq:fkexpansion}).  Now consider a specific $H^2R^3$ term with identical kinematical structure but tree-level coefficient $A_T$ and one-loop coefficient $A_L$.  Equating coefficients with the above then gives
\begin{equation}
a_0=\fft{A_T+3A_L}4,\qquad a_1=-\fft{3(A_T-A_L)}8.
\label{eq:ATAL}
\end{equation}
This now allows us to write the quintic terms in the middle two lines of (\ref{eq:IIB}) as
\begin{align}
\mathcal L_{H^2R^3}&=\fft{\alpha'^3}{3\cdot2^{12}}\sqrt{-g}\Bigl[\epsilon_9\epsilon_9\left(-\ft7{24}
f_0(\tau,\bar\tau)|G|^2-\ft1{16}(f_1(\tau,\bar\tau)G^2+c.c.)\right)R^3\nn\\
&\kern7em+t_8t_8\left(-\ft12
f_0(\tau,\bar\tau)|G|^2+\ft34(f_1(\tau,\bar\tau)G^2+c.c.)\right)R^3\nn\\
&\kern7em+2\cdot4!\sum_id_i\left(f_0(\tau,\bar\tau)G^{\mu\nu\lambda}\bar G^{\rho\sigma\zeta}-\ft32\left(f_1(\tau,\bar\tau)G^{\mu\nu\lambda}G^{\rho\sigma\zeta}+c.c.\right)\right)\tilde Q^i_{\mu\nu\lambda\rho\sigma\zeta}+\cdots\Bigr].
\label{eq:quintic}
\end{align}
This is fully constrained by knowledge of the tree and loop coefficients along with $SL(2,\mathbb Z)$ invariance.  Note, however, that higher order couplings in the NSNS sector which can be extended to a linear combination of terms with more than two possible $U(1)$ charge assignments will not have a unique $SL(2,\mathbb Z)$ completion without additional input.

Finally, we consider the last line in \eqref{eq:IIB} containing the CP-odd contribution to the action. Note that the part of  $X_8(\hat\O_+) \sim [\tr R(\hat\Omega_+)^4-\ft14(\tr R(\hat\Omega_+)^2)^2]$ odd in $B_2$ can be written as $X_8(\hat\O_+) - X_8(\hat\O_-)$ and as a difference of the same characteristic classes with different connections is exact. Hence the CP-odd term can be written as $H \wedge X_7 (\O, H)$.  The natural extension of this term to include the RR sector is then schematically
\begin{equation}
\mathcal L_{\textrm{CP-odd}}=\fft{(2\pi)^2}{3\cdot2^6}\alpha'^3e^{\fft12\phi}\left(G\wedge X_7(\Omega,G,\bar G)+c.c.\right).
\end{equation}
The presence of the factor $e^{\fft12\phi}$ is consistent with this being a one-loop term, and demonstrates it necessarily transforms non-trivially under $SL(2,\mathbb Z)$.  For the five-point function, the same decomposition into $U(1)$ preserving and violating terms as done in (\ref{eq:a0a1}) and (\ref{eq:ATAL}) applies.  This time, however, $A_T=0$ since there is no corresponding CP-odd $H^2R^3$ term at tree level.  The $SL(2,\mathbb Z)$ completion is then
\begin{equation}
\mathcal L_{\textrm{CP-odd}}=\fft{3\alpha'^3}{2^8}\left[G\wedge\left(f_0(\tau,\bar\tau)X_7^{(1)}(\Omega,\bar G)+f_1(\tau,\bar\tau)X_7^{(1)}(\Omega, G)+\cdots\right)+c.c.\right],
\label{eq:CPoG}
\end{equation}
where $X_7^{(1)}(\Omega,G)$ denotes the contribution to $X_7(\Omega,G,\bar G)$ linear in $G$.  Just as in the CP-even sector, higher-order non-linear terms cannot be fully pinned down without further input on the $U(1)$ charge assignments.

An interesting feature of the CP-odd sector is that while there is no tree-level CP-odd term in the NSNS sector, the $SL(2;\mathbb Z)$ completion, (\ref{eq:CPoG}), requires a tree-level CP-odd coupling of the form $F_3\wedge X_7(\Omega,F_3)$.  The presence of such a tree-level term involving $\epsilon_{10}$ from a string world-sheet point of view may appear as a surprise as it cannot arise in amplitudes containing only NSNS emission vertex operators.  However, such a term can exist in the RR sector since $\epsilon_{10}$ is naturally present when working with chiral ten-dimensional spinors.  It would be interesting to verify this coupling directly from a five-point tree amplitude involving two RR fields.

Since the non-holomorphic Eisenstein series $f_k(\tau,\bar\tau)$ expands into tree-level, one-loop and set of non-perturbative terms with known coefficients, once the tree-level and one-loop couplings are determined, the non-perturbative couplings can then be read off from the large $\tau_2$ expansion, (\ref{eq:fkexpansion}).  In this sense, the quintic effective action (\ref{eq:quintic}) predicts a new set of non-perturbative couplings with non-standard kinematics.  It would be interesting to study this further and see what implications it may have for D-instantons \cite{Green:1997tv, Green:1997di} or other non-perturbative objects in type IIB theory.  Even at the perturbative level, (\ref{eq:quintic}) makes predictions for RR sector couplings without any explicit computation of RR amplitudes.  This ability to make predictions in the RR sector also extends to the use of $O(5,21)$ duality in the K3 compactification of IIB theory to $\mathcal N=(2,0)$ theory in six dimensions.

\section{Discussion}
\label{sec:disc}

We have made progress towards the determination of the eight-derivative quintic effective action of type II strings by computing the tree-level $H^2R^3$ couplings.  However, even in the NSNS sector, this calculation is not complete.  What is missing are possible $H^2(\nabla H)^2R$ terms beyond those that originate from $R(\Omega_+)^4$ as well as couplings involving the dilaton.  In six-dimensional tests of our couplings, where two gravitons are internal, we could reduce the tree-level action to four-point contact terms in six dimensions that involve the dilaton, (\ref{eq:tree4}).  These terms are expected to cancel based on supersymmetry, but their cancellation would clearly require new six-point contact terms in ten-dimensions, including those containing the dilaton. However we used on-shell manipulations such as those in Appendix~\ref{sec:Hsimp} that allowed us to push some ostensibly three-point terms to the four-point level. This was done in six dimensions, and is not bound to work in ten. Hence our tests do not rule out the potential appearance of dilaton couplings already in the ten-dimensional five-point contact terms.

In principle, it would not be difficult to extend the calculation of section~\ref{sec:stam} to recover the $H^2(\nabla H)^2R$ contact terms by considering four antisymmetric tensor and one graviton scattering, although we would have to consider a greater number of invariants when constructing the effective action.  Likewise, the dilaton can be included as well, although we would have to explore a wider number of amplitudes involving from one up to five external dilatons along with an even number of antisymmetric tensors and some number of remaining gravitons.  Nevertheless, given the interesting differences between tree-level and one-loop, and the fact that the dilaton is closely related to loop counting, it would be important to complete our results with the dilaton.

In addition, there is more to the type II effective action than just the NSNS sector.  Pure spinor techniques have been applied to open string fermion amplitudes, and in principle they can be combined to yield the corresponding RR amplitudes.  This was done at the level of the quartic action \cite{Policastro:2006vt}, and we expect it should be possible to do the same for the quintic action.  Of course, the usual issue of subtracting out the underlying pole terms from the lower-point functions will still have to be done.

Even without a complete quintic effective action, we were able to test the new couplings in some simple compactifications.  One interesting observation arises from the compactification of type II strings on K3.  Based on heterotic/IIA duality, we can obtain the quartic one-loop action of IIA in six dimensions \cite{Liu:2013dna} and map it to IIB in six dimensions.  Reconciling this action with six-dimensional $(2,0)$ supersymmetry then required a non-trivial interplay between the CP-even and CP-odd sectors of the theory.  When lifted to ten dimensions, this provides strong evidence that IIB theory indeed has a one-loop CP-odd term, although it is cohomologically trivial and hence does not represent any anomaly but is nevertheless required by supersymmetry.

Finally, perhaps one of the most intriguing outcomes of this investigation into the non-linear completion of $R^4$ is a better appreciation on the intricate $SL(2,\mathbb Z)$ structure of the eight-derivative IIB couplings.  While the picture is still incomplete, the appearance of different numerical factors and kinematical structures between tree-level and one-loop gives rise to a rather complicated set of terms that simplify when reduced to the purely gravitational sector.  In addition, $SL(2,\mathbb Z)$ invariance allows us to deduce some information on the RR sector as well as non-perturbative information, even without any direct computations.  These hints of various structures suggest that we look for some sort of underlying generalised geometrical description of the higher-derivative couplings or perhaps even some larger hidden symmetries of string theory.


\subsubsection*{Acknowledgments}

Useful discussions with  M. Cicoli, A. Coimbra, H. Elvang, T. Grimm, S. Katmadas, B. Pioline, F. Quevedo, R. Savelli, A. Schachner, S. Steiberger, R. Valandro and Y. Wang are gratefully acknowledged. We are grateful to R. Savelli and A. Schachner for correcting an initial sign error in (\ref{eq:a0a1}).  This work was supported in part by the US Department of Energy under grant DE-SC0007859 (JTL) and by ERC grants 772408-Stringlandscape and 787320-QBH Structure (RM).


\appendix


\section{The \texorpdfstring{$R^3$}{R3} basis tensors}
\label{app:R3}

As shown in section~\ref{sec:tree5}, there are 24 independent Lorentz invariants that can be formed out of $H^2R^3$.  We hence introduce a 24-dimensional basis labeled by the number of indices contracted between $H^2$ and $R^3$.  The $R^3$ invariants themselves are denoted $\{\tilde S^i,\tilde W_{\mu\nu}^i,\tilde X_{\mu\nu\rho\sigma}^i,\tilde Q_{\mu\nu\lambda\rho\sigma\zeta}^i\}$, where the tilde is introduced as a reminder that these tensors (except for the singlets $S^i$) transform reducibly under $SO(1,9)$.

There are two singlet combinations
\begin{align}
\tilde S^1&=R_{ab}{}^{cd}R_{cd}{}^{ef}R_{ef}{}^{ab},\nn\\
\tilde S^2&=R^c{}_{ab}{}^dR^e{}_{cd}{}^fR^a{}_{ef}{}^b,
\end{align}
and three symmetric two-index combinations
\begin{align}
\tilde W^1_{\mu\nu}&=R_{\mu a\nu b}R^a{}_{cde}R^{bcde},\nn\\
\tilde W^2_{\mu\nu}&=R_{\mu eab}R_\nu{}^{ecd}R^{ab}{}_{cd},\nn\\
\tilde W^3_{\mu\nu}&=R_{\mu aeb}R_\nu{}^{ced}R^a{}_c{}^b{}_d.
\end{align}
For the four-index combinations, we take a combination of $[\mu\nu][\rho\sigma]$ and $[\mu\nu\rho\sigma]$ tensor structures (along with trace terms which we do not project out).  There are a total of eleven independent terms
\begin{align}
\tilde X^1_{\mu\nu\rho\sigma}&=R_{\mu\nu\rho\sigma}R_{abcd}R^{abcd},\nn\\
\tilde X^2_{\mu\nu\rho\sigma}&=R_{\mu\nu\rho a}R_{\sigma bcd}R^{abcd},\nn\\
\tilde X^3_{\mu\nu\rho\sigma}&=R_{\mu\nu ab}R_{\rho\sigma cd}R^{abcd},\nn\\
\tilde X^4_{\mu\nu\rho\sigma}&=R_{\mu\rho ab}R_{\nu\sigma cd}R^{abcd},\nn\\
\tilde X^5_{\mu\nu\rho\sigma}&=R_{\mu a\rho c}R_{\nu b\sigma d}R^{abcd},\nn\\
\tilde X^6_{\mu\nu\rho\sigma}&=R_{\mu\rho ab}R_\nu{}^a{}_{cd}R_\sigma{}^{bcd},\nn\\
\tilde X^7_{\mu\nu\rho\sigma}&=R_{\mu a\rho b}R_\nu{}^a{}_{cd}R_\sigma{}^{bcd},\nn\\
\tilde X^8_{\mu\nu\rho\sigma}&=R_{\mu\rho ab}R_{\nu c}{}^a{}_dR_\sigma{}^{cbd},\nn\\
\tilde X^9_{\mu\nu\rho\sigma}&=R_{\mu a\rho b}R_{\nu c}{}^a{}_dR_\sigma{}^{cbd},\nn\\
\tilde X^{10}_{\mu\nu\rho\sigma}&=R_{\mu\nu}{}^{ab}R_{\rho acd}R_{\sigma b}{}^{cd},\nn\\
\tilde X^{11}_{\mu\nu\rho\sigma}&=R_{\mu\nu ab}R_{\rho c}{}^a{}_dR_\sigma{}^{cbd}.
\end{align}
Finally, we have eight CP-even six-index combinations from $[\mu\nu\lambda][\alpha\beta\gamma]$ and $[\mu\nu\lambda\alpha\beta]\gamma$
\begin{align}
\label{eq:Qi}
\tilde Q^1_{\mu\nu\lambda\alpha\beta\gamma}&=R_{\mu\alpha a}{}^bR_{\nu\beta b}{}^cR_{\lambda c\gamma}{}^a,\nn\\
\tilde Q^2_{\mu\nu\lambda\alpha\beta\gamma}&=R_{\mu\nu a}{}^bR_{\alpha\beta b}{}^cR_{\lambda c\gamma}{}^a,\nn\\
\tilde Q^3_{\mu\nu\lambda\alpha\beta\gamma}&=R_{\mu\nu a}{}^bR_{\lambda\alpha b}{}^cR_{\beta c\gamma}{}^a,\nn\\
\tilde Q^4_{\mu\nu\lambda\alpha\beta\gamma}&=R_{\mu a\alpha}{}^bR_{\nu b\beta}{}^cR_{\lambda c\gamma}{}^a,\nn\\
\tilde Q^5_{\mu\nu\lambda\alpha\beta\gamma}&=R_{\mu abc}R_{\nu\alpha}{}^{bc}R_{\lambda\beta\gamma}{}^a,\nn\\
\tilde Q^6_{\mu\nu\lambda\alpha\beta\gamma}&=R_{\mu abc}R_{\alpha\beta}{}^{bc}R_{\nu\lambda\gamma}{}^a,\nn\\
\tilde Q^7_{\mu\nu\lambda\alpha\beta\gamma}&=R_{\mu abc}R_\nu{}^a{}_\alpha{}^cR_{\lambda\beta\gamma}{}^b,\nn\\
\tilde Q^8_{\mu\nu\lambda\alpha\beta\gamma}&=R_{\mu\nu\alpha\beta}R_{\lambda abc}R_\gamma{}^{abc}.
\end{align}

\section{Simplification of \texorpdfstring{$H^{(-)\,2}\nabla\nabla\phi$ }{HHddphi}}
\label{sec:Hsimp}

Here we prove the on-shell identity
\begin{equation}
H^{(-)\,2}_{\mu\nu}\nabla^\mu\nabla^\nu\phi=-2H^{(-)\,2}_{\mu\nu}\partial^\mu\phi\partial^\nu\phi-\ft16H^2\partial\phi^2.
\label{eq:HHddpid}
\end{equation}
We start with the observation
\begin{equation}
H^{(-)\,2}_{\mu\nu}\nabla^\mu\nabla^\nu\phi=\nabla^\mu(H^{(-)\,2}_{\mu\nu}\partial^\nu\phi)-(\nabla^\mu H^{(-)}_\mu{}^{\alpha\beta})H^{(-)}_{\nu\alpha\beta}\partial^\nu\phi-H^{(-)\,\mu}{}_{\alpha\beta}\nabla_\mu H^{(-)}_{\nu\alpha\beta}\partial^\nu\phi.
\end{equation}
Although we need to work with the anti-self dual component of $H$, we can use a combination of Bianchi and equations of motion to manipulate these expressions.  For the last term, we start by replacing $H^{(-)}\to-*H^{(-)}$ for both of the $H^{(-)}$'s
\begin{align}
H^{(-)\,\mu}{}_{\alpha\beta}\nabla_\mu H^{(-)}_{\nu\alpha\beta}&=\fft1{3!^2}\epsilon^\mu{}_{\alpha\beta}{}^{\lambda_1\lambda_2\lambda_3}\epsilon_{\nu\alpha\beta\sigma_1\sigma_2\sigma_3}H^{(-)}_{\lambda_1\lambda_2\lambda_3}\nabla_\mu H^{(-)\,\sigma_1\sigma_2\sigma_3}\nn\\
&=-\fft1{18}\delta^{\mu\lambda_1\lambda_2\lambda_3}_{\nu\sigma_1\sigma_2\sigma_3}H^{(-)}_{\lambda_1\lambda_2\lambda_3}\nabla_\mu H^{(-)\,\sigma_1\sigma_2\sigma_3}\nn\\
&=-\fft1{18}(\delta^\mu_\nu\delta^{\lambda_1\lambda_2\lambda_3}_{\sigma_1\sigma_2\sigma_3}-3\delta_\nu^{[\lambda_1|}\delta^{\mu|\lambda_2\lambda_3]}_{\sigma_1\sigma_2\sigma_3})H^{(-)}_{\lambda_1\lambda_2\lambda_3}\nabla_\mu H^{(-)\,\sigma_1\sigma_2\sigma_3}\nn\\
&=-\fft13\delta^\mu_\nu H^{(-)}_{\lambda_1\lambda_2\lambda_3}\nabla_\mu H^{(-)\,\lambda_1\lambda_2\lambda_3}+H^{(-)}_{\nu\lambda_2\lambda_3}\nabla_\mu H^{(-)\,\mu\lambda_2\lambda_3}\nn\\
&=-\fft16\nabla_\nu(H^{(-)\,2})+H^{(-)}_{\nu\lambda_2\lambda_3}\nabla_\mu H^{(-)\,\mu\lambda_2\lambda_3}.
\end{align}
The first term vanishes kinematically since $H^{(-)}\cdot H^{(-)}=0$ for anti-self dual $H^{(-)}$.  As a result, we find
\begin{equation}
H^{(-)\,2}_{\mu\nu}\nabla^\mu\nabla^\nu\phi=-2(\nabla^\mu H^{(-)}_\mu{}^{\alpha\beta})H^{(-)}_{\nu\alpha\beta}\partial^\nu\phi,
\label{eq:Hnnp}
\end{equation}
where we have dropped the total derivative (at one loop).  We now write $H^{(-)}$ in terms of $H$ and its dual
\begin{equation}
H^{(-)}_{\mu\alpha\beta}=\fft12\left(H_{\mu\alpha\beta}-\fft1{3!}\epsilon_{\mu\alpha\beta}{}^{\rho\delta\sigma}H_{\rho\delta\sigma}\right).
\end{equation}
Taking a divergence then gives
\begin{equation}
\nabla^\mu H^{(-)}_{\mu\alpha\beta}=\fft12\left(\nabla^\mu H_{\mu\alpha\beta}-\fft1{3!}\epsilon_{\alpha\beta}{}^{\mu\rho\delta\sigma}\partial_\mu H_{\rho\delta\sigma}\right).
\end{equation}
The second term vanishes by Bianchi, $dH=0$.  For the first term, we use the equation of motion (\ref{eq:eom0}) to arrive at
\begin{equation}
\nabla^\mu H^{(-)}_{\mu\alpha\beta}=H_{\mu\alpha\beta}\partial^\mu\phi.
\end{equation}
Inserting this into (\ref{eq:Hnnp}) then gives
\begin{equation}
H^{(-)\,2}_{\mu\nu}\nabla^\mu\nabla^\nu\phi=-2H_{\mu\alpha\beta}H^{(-)}_\nu{}^{\alpha\beta}\partial^\mu\phi\partial^\nu\phi.
\label{eq:Hmsimp}
\end{equation}
Finally, we break $H_{\mu\alpha\beta}$ into its self dual and anti-self dual components and make use of the identity (\ref{eq:Hpmmn})
\begin{equation}
H_{\mu\alpha\beta}H^{(-)}_\nu{}^{\alpha\beta}\partial^\mu\phi\partial^\nu\phi=H^{(-)\,2}_{\mu\nu}\partial^\mu\phi\partial^\nu\phi+H^{(+)}_{\mu\alpha\beta}H^{(-)}_\nu{}^{\alpha\beta}\partial^\mu\phi\partial^\nu\phi=H^{(-)\,2}_{\mu\nu}\partial^\mu\phi\partial^\nu\phi+\ft1{12}H^2\partial\phi^2.
\end{equation}
Combining this with (\ref{eq:Hmsimp}) then proves the identity (\ref{eq:HHddpid}).




\providecommand{\href}[2]{#2}\begingroup\raggedright\endgroup


\end{document}